\documentclass[useAMS,usenatbib]{mn2e}
\usepackage{graphicx}

\newcommand{\lre}{$\log \rm R_{\rm e}$}
\newcommand{\re}{$\rm R_{\rm e}$}

\newcommand{\sn}{$n$}
\newcommand{\mie}{$\langle\mu\rangle_{\rm e}$}

\newcommand{\papdata}{Paper I}

\newcommand{\nablat}{$\nabla_t$}
\newcommand{\nablati}{$\nabla_{t,i}$}

\newcommand{\nablaz}{$\nabla_Z$}
\newcommand{\nablazi}{$\nabla_{Z,i}$}
\newcommand{\nablazo}{$\nabla_{Z,o}$}
\newcommand{\dlt}{$\Delta_t$}
\newcommand{\simlt}{\lower.5ex\hbox{$\; \buildrel < \over \sim \;$}}
\newcommand{\simgt}{\lower.5ex\hbox{$\; \buildrel > \over \sim \;$}}

\begin{document}
\title[Stellar populations of massive ETGs out to $\rm R\simlt 8\,\rm R_e$.]
{SPIDER VII -- Revealing the Stellar Population Content of Massive Early-type Galaxies out to $8\,\rm R_e$.}
\author[F. La Barbera et al.]{F.  La  Barbera$^{(1)}$, I.  Ferreras$^{(2)}$, R.~R. de Carvalho$^{(3)}$, 
G. Bruzual$^{(4)}$, S. Charlot$^{(5,6)}$,  \newauthor
A. Pasquali$^{(7)}$,
E. Merlin$^{(8)}$\\
$^1$INAF -- Osservatorio Astronomico di Capodimonte, Napoli, Italy \\
$^2$MSSL, University College London, Holmbury St Mary, Dorking, Surrey RH5 6NT\\
$^3$Instituto Nacional de Pesquisas Espaciais/MCT, S. J. dos Campos, Brazil\\
$^4$Centro de Radioastronom\'\i a y Astrof\'\i sica, UNAM, Campus Morelia, M\'exico\\
$^5$UPMC, UMR7095, Institut d'Astrophysique de Paris, F-75014 Paris, France\\
$^6$CNRS, UMR7095, Institut d'Astrophysique de Paris, F-75014 Paris, France\\
$^7$Astronomisches Rechen Institut, Zentrum f\"ur Astronomie der Universit\"at Heidelberg, M\"onchhofstr. 12--14, 69120 Heidelberg, Germany\\
$^8$INAF -- Osservatorio Astronomico di Padova, Padova, Italy\\
}

\date{Accepted 2012 August 01. Received 2012 July 06; in original form 2012 May 26}
\pagerange{\pageref{firstpage}--\pageref{lastpage}} \pubyear{2012}
\maketitle
\label{firstpage}

\begin{abstract}
Radial trends of stellar populations in galaxies provide a valuable
tool to understand the mechanisms of galaxy growth. In this paper, we present 
the first comprehensive analysis of optical--optical and optical--NIR colours, 
as a function of galaxy mass, out to the halo region ($8\,\rm R_e$) of early-type 
galaxies (ETGs). We select a sample of 674 massive ETGs ($M_\star\simgt
3\times 10^{10}M_\odot$) from the SDSS-based SPIDER survey.  By 
comparing with a large range of population
synthesis models, we derive robust constraints on the radial trends in
age and metallicity. Metallicity is unambiguously found to decrease
outwards, with a measurable steepening of the slope in the outer
regions ($\rm R_e<R<8\,R_e$). The gradients in stellar age are found to be more sensitive to
the models used, but in general, the outer regions of ETGs feature
older populations compared to the cores. This trend is strongest for
the most massive galaxies in our sample ($M_\star\simgt
10^{11}M_\odot$). Furthermore, when segregating with respect to large
scale environment, the age gradient is more significant in ETGs
residing in higher density regions. These results shed light on the
processes leading from the formation of the central core to the growth
of the stellar envelope of massive galaxies. The fact that the
populations in the outer regions are older and more metal-poor
than in the core suggests a process whereby the envelope of massive
galaxies is made up of accreted small satellites (i.e. minor mergers)
whose stars were born during the first stages of galaxy formation.
\end{abstract}

\begin{keywords}
galaxies: fundamental parameters -- galaxies: elliptical and
lenticular, cD -- galaxies: evolution -- galaxies: formation --
galaxies: stellar content -- galaxies: groups: general
\end{keywords}

\section{Introduction}

The link between chemical and dynamical evolution in early-type
galaxies (hereafter ETGs) has been studied for quite a long time
through systematic measurements of colour gradients, which may lead us
to understanding how the variations in the age and metallicity of the
underlying stellar populations evolve as star formation proceeds
through the history of the system
\citep{devac61,davies87,PVJ90,ferr05,LdC:09}.  The theoretical models
suggest that metallicity and age gradients arise naturally from the
processes leading to galaxy formation, like in the ``in-situ''
collapse of a proto-galactic gas cloud \citep{carlberg84}.  The first
rendition of this formation scenario, also known as monolithic
collapse \citep{els62,lar74,lar75}, predicted steep metallicity
gradients, while, a revised monolithic model for the formation of ETGs
led to shallower metallicity gradients
\citep[e.g.][]{pip08,pip10}. Simulations \citep[see e.g.][]{koba04}
indicate that galaxy mergers will weaken the metallicity gradient, and
lead to significant levels of star formation in the galaxy
centre. { More recently, cosmological simulations have brought to our
attention the importance of cold accretion, as a viable way to form galaxies  ``in-situ'', resembling the monolithic formation scenario~\citep[e.g.][]{keres05}.}
It is still a matter of debate whether
the metallicity gradient we measure today is a product of an old
stellar population -- reflecting mainly the initial conditions when
the system collapsed -- or results from more recent star formation
episodes. Therefore, systematic measurements of colour gradients are
of crucial importance to distinguish among different formation models
of early-type systems. A proper constraint on the gradients of the
properties of the underlying stellar populations will help understand
the mass assembly process in ETGs.

In the past, spectroscopic indices were used to measure metallicity in
ETGs, revealing the existence of radial gradients ranging from $-$0.1
to $-$0.3~dex per decade \citep{cmc93,davies93,mehlert03}. The number
of ETGs for which the radial dependence of age and metallicity extends
beyond one effective radius is small for today standards, albeit
growing. This is due to the difficulty in obtaining spectroscopic
measurements at surface brightness below the background sky
level. However, the availability of multi-waveband samples for a vast
number of sources remedy this situation, allowing us to estimate
colour gradients up to several effective radii at low to moderate
redshifts \citep{wu05,TalvanDokkum:11}, and open the possibility of
studying radial gradients at z$>$1 \citep{Gargiulo:11, GUo:11}.  An
important caveat is that colours have to be modelled by a stellar
population synthesis code. This approach has been used in the past for
small samples of ETGs \citep[e.g.][]{PVJ90,SMG00}, while in the recent
years, the study of large galaxy samples has mostly focused on the
optical regime \citep[e.g.][]{RBH:10,Tor:10, GP11}, where one has to
contend with the age-metallicity-extinction degeneracy \citep{wo94}.
\citet{SE94}, and more recently~\citet{Roediger:11}, have found
evidence for significant age gradients in ellipticals, while
independent works \citep[e.g.][]{HI01,mehlert03}, found negligible age
gradients, showing that the interpretation of colour gradients is a
difficult task.

{  Another  aspect of  studying  colour  gradients  in ETGs  is  to
  understand  how  this  quantity  relates to  other  observables  and
  relationships so that we can  build up a consistent galaxy formation
  and evolution framework. For  instance, although} ETGs obey specific
scaling relations  (e.g. Fundamental  Plane), they may  well represent
different  families according  to the  way mass  was  assembled during
their  history  and  the  characterization  of how  mergers  may  have
established   the  mass   assembly  is   of   unquestionable  interest
\citep[e.g.][]{Greene:12}. As  a first approximation,  we would expect
mergers  to  significantly  affect  the well-known  scaling  relations
involving   central  velocity   dispersion   and  stellar   population
properties, unless mass is  accreted at very large radii.  Theoretical
and observational work has shown  that the presence of tidal debris in
the outskirts  of ETGs may provide  an essential piece  of evidence to
distinguish among different assembly scenarios \citep[e.g.][]{Duc:11}.
So far, most of the stellar population work was carried out within the
central  regions   and  mainly  restricted  to   the  optical  regime.
Extending  the analysis  to the  outer regions  with a  combination of
optical and  near-infrared data is  a key factor to  better comprehend
how galaxies form and assemble their baryons.

{ In this paper we  combine optical (Sloan Digital Sky Survey--Data
  Release  6,  SDSS--DR6;~\citealt{ade08})  and  near-infrared  (UKIRT
  Infrared    Deep   Sky    Survey--Data   Release    4,   UKIDSS-DR4;
  \citealt{Law07}) } data to study colour gradients in a sample of 674
ETGs located in different environments.  The data were analysed with a
dedicated pipeline  that uses the package  2DPHOT \citep{LBdC08}.  The
high  quality image  of both  optical and  NIR surveys  enabled  us to
determine reliable  colour gradients out to eight  effective radii, as
shown below. All systematic effects were carefully taken into account,
including background subtraction; stacking  of the colour profiles and
the effect of the tails of the  PSFs in the redder bands, all of which
are crucial  for an accurate measurement  of the colours  in the outer
regions.  The  results presented here confirm  previous findings about
the metallicity gradients,  and show that age gradients  may not be as
negligible as some papers have reported in the past.

The layout of the paper is as follows. In Sec.~\ref{sec:samples} we
present our sample of ETGs. Sec.~\ref{sec:profiles} deals with the
methodology followed to determine the colour profiles, and discusses
possible sources of systematics. In Sec.~\ref{sec:col_prof_mass} we
present the median-stacked colour profiles of ETGs, from $g-r$ through
$g-K$, out to a maximum galactocentric distance of $8\,\rm R_e$, for
different galaxy mass bins. Sec.~\ref{sec:fitting} describes the
fitting of the observed profiles with theoretical models based on
different, state-of-the-art, stellar population synthesis codes. In
Sec.~\ref{sec:SP_Age_Z} we show the results of the stellar population
fits, i.e. how age and metallicity are found to vary from the centre
to the external regions of ETGs.  We also discuss the contribution of
internal reddening. Sec.~\ref{sec:environment} shows how the inferred
age and metallicity profiles are affected by galaxy environment, while
in Sec.~\ref{sec:summary} we summarize the main findings of this
paper.  Throughout the paper, we adopt a standard $\Lambda$CDM
cosmology with $\rm H_0 \! = \!  75\, km \, s^{-1} \, Mpc^{-1}$,
$\Omega_{\rm m} \!  = \!  0.3$, and $\Omega_{\Lambda} \! = \! 0.7$.

\begin{figure*}
\begin{center}
\includegraphics[height=140mm]{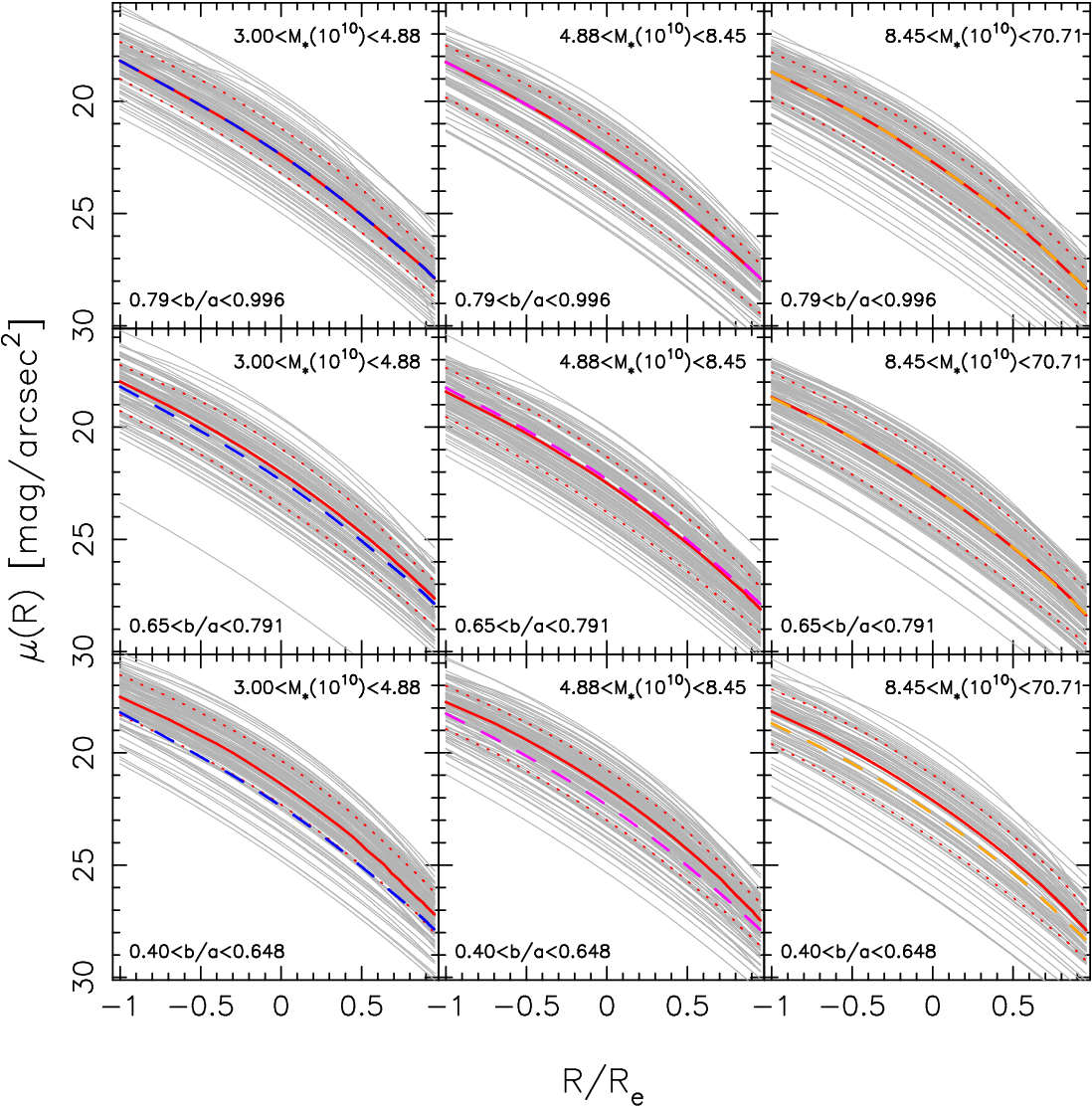}
\end{center}
\caption{Comparison of the median surface brightness profiles of ETGs
  in different bins of stellar mass, $M_{\star}$ (increasing from left
  to right), and axis ratio, $b/a$ (decreasing from top to
  bottom). For each panel (i.e. a given bin of $M_{\star}$ and $b/a$),
  black curves plot the $r$-band surface brightness profiles of all galaxies
  in that bin. The profiles are shown as a function of the
  normalized galactocentric distance $\rm R/R_e$, where $\rm R_e$ is
  the $r$-band effective radius.  For each panel, the solid red curve is
  obtained by median-stacking all the single profiles, with dotted
  curves marking the $\pm 1\sigma$ scatter around the median
  profiles. For each bin of $M_{\star}$, the median profile that
  corresponds to highest $b/a$ (top panel) is repeated in all panels
  (from top to bottom) as a coloured dashed curve (blue, magenta, and
  orange for left, middle, and right panels, respectively).  }
\label{fig:profs}
\end{figure*}

\section{The sample}
\label{sec:samples}

The present study is based on  a set of $674$ ETGs, extracted from the
SPIDER survey~\citep[hereafter  Paper I]{PaperI}. This  sample selects
those galaxies with  the best quality of the  parameters that describe
their  surface  brightness  distributions.   {  The  SPIDER  sample
  consists of $39,993$ luminous ETGs, in the redshift range of 0.05 to
  0.095,  with $M_{r}{<}-20$,  where $M_{r}$  is the  k-corrected SDSS
  Petrosian magnitude  in r-band. All galaxies  have $griz$ photometry
  and  spectroscopy from  SDSS-DR6,  while $5,080$  ETGs  } also  have
photometry in  the $YJHK$ wavebands from the  UKIDSS-Large Area Survey
(see~\papdata).  In all wavebands, the structural parameters, i.e. the
effective  radius, \re,  mean surface  brightness within  that radius,
\mie, and S\'ersic index,  \sn, have been homogeneously measured using
the  software  2DPHOT~\citep{LBdC08}, by  fitting  galaxy images  with
two-dimensional seeing-convolved S\'ersic models. Total magnitudes are
computed  from \re\  and  \mie\  in each  filter.  Stellar masses  are
derived   by  fitting   synthetic  stellar   population   (SP)  models
from~\citet{BrC03},  with a  variety of  star formation  histories and
metallicities,  to  the  optical+NIR  photometry, using  the  software
LePhare~\citep{Ilbert:06}, assuming  a \citet{Cardelli} extinction law
and  Chabrier IMF  (see~\citealt[hereafter  Paper V]{Swindle:11},  for
details on the estimate of $M_\star$).
 

\medskip

To perform an accurate stacking of the surface brightness profiles
(seeing de-convolved) as a function of galactocentric distance (see
Sec.~\ref{sec:profiles}), we select ETGs according to the following
criteria.

\begin{description}
 \item[-] We select only galaxies whose light distribution is well
   fitted by a S\'ersic model, i.e. with $\chi^2<2$ in all wavebands,
   where $\chi^2$ is the rms of residuals between the galaxy image in
   a given band and the corresponding best-fitting two-dimensional
   S\'ersic model. We also remove galaxies with large uncertainties
   ($>0.5$~dex) in \lre, leading to a sample of $4,546$ (out of $5,080$) ETGs. This
   sample is the same one used in~\citet[hereafter Paper IV]{PaperIV},
   where the inner ($<1\,\rm R_e$) colour gradients were analyzed,
   along with their dependence on mass and stellar population
   properties (i.e. age, metallicity, and [$\alpha$/Fe]).
 \item[-] We select galaxies in a narrow redshift range, from $0.05$
   to $0.07$, which corresponds to those with the best S/N ratio,
   minimizing the effect of k- and evolutionary corrections varying
   within the sample and as a function of galactocentric distance.
   This results in $1,255$ ETGs.
 \item[-] Galaxies  with stellar  mass less than  $M_\star =  3 \times
   10^{10} M_{\odot}$ are removed,  leading to $1,043$ ETGs.  { The
     reason for this selection is the following. At the upper redshift
     limit  of $z=0.095$,  the magnitude  limit of  the  SPIDER sample
     ($M_r  \sim -20$;  see above),  corresponds approximately  to the
     magnitude limit below  which SDSS spectroscopy becomes incomplete
     (i.e.  a Petrosian magnitude of $m_r \sim 17.8$).  This makes the
     SPIDER  sample incomplete  at  a stellar  mass  of $M_\star  \sim
     10^{10}  M_{\odot}$  (corresponding  to  $M_r  \sim  -20$).   The
     selection limit of $M_\star = 3 \times 10^{10} M_{\odot}$ ensures
     that the sample  of ETGs used for the  present analysis is $90\%$
     complete with  respect to  stellar mass, this  completeness level
     being  estimated  from  the  $M_\star$  versus  r-band  Petrosian
     magnitude plot (see, e.g., sec.~3.2 of Paper IV)}.
 \item[-] Finally we remove galaxies with an axis ratio, $b/a \le
   0.65$. The reason for this selection is illustrated in
   Fig~\ref{fig:profs}, that compares the r-band median-stacked
   surface brightness profiles (solid red curves) of the $1,043$ ETGs
   at $0.05 \le z \le 0.07 $ (see above) for three different bins of
   $b/a$ (from top to bottom) and $M_{\star}$ (from left to right).
   For a given bin of $M_{\star}$, the median profile corresponding to
   the highest $b/a$ is shown in all the panels as a dashed curve. At
   all masses, the profile does not change significantly (within a few
   percent) for $b/a\simgt 0.65$, justifying our cut based on
   $b/a$. As argued by \citet{HydeBernardi:09}, ETGs with $b/a \simlt
   0.6$ may indeed represent a different population of objects
   (e.g. rotationally supported systems).
\end{description}

Our final sample of high-quality data comprises $674$ ETGs, and we
split this sample in three bins of stellar mass, keeping the number of
galaxies per bin fixed. Hereafter, we refer to galaxies in the three
bins as low- ($3.00 < M_{\star} [10^{10} M_{\odot}] <5.29$),
intermediate- ($5.29 < M_{\star} [10^{10} M_{\odot}] < 9.42$), and
high- ($9.42 < M_{\star} [10^{10} M_{\odot}] <70.71$) mass,
respectively. In each bin of galaxy mass, we also split galaxies
according to the environment where they reside.  To this effect, we
use the friends-of-friends (FoF) group catalogue, as described in
\citet{Berlind:06}. We use an updated catalogue based on SDSS-DR7
-- rather than DR3, used for the original \citet{Berlind:06}
dataset. The catalogue, that comprises $10,124$ systems, is subject
to a virial analysis \citep[see][and references therein]{lop09a},
resulting in a compilation of $8,083$ groups with well measured
properties \citep[see][hereafter Paper III, for details]{PaperIII}.
ETGs in the SPIDER sample are then classified as either (i) group
galaxies, i.e. those having group membership according to the virial
analysis ($\sim46\%$); (ii) field galaxies ($\sim 33 \%$),
i.e. objects located more than five virial radii away from any group
initially detected by the FoF algorithm; and (iii) unclassified ($\sim
21 \%$), i.e. objects in neither of the other two classes. Out of the
selected sample of $674$ ETGs analyzed in the present work, 336 (205)
objects are group (field) galaxies. The group galaxies reside in systems 
with average velocity dispersion (virial mass) of 
$\sim 240 \rm km \, s^{-1}$ ($\sim 8 \cdot 10^{13} M_\odot$).
These subsamples are used to
investigate the dependence of stellar population profiles in ETGs on
 environment (Sec.~\ref{sec:environment}).

\section{Derivation of colour profiles}
\label{sec:profiles}
\subsection{Methodology}
For a given waveband $X$, with $X=grizYJHK$, the surface brightness
profile of each ETG, $\mu_X(R)$, is modelled by the S\'ersic law:
\begin{equation}
\mu_X(R)=\mu_{0,X} + 1.0857 \cdot b_{n,X} \left(\frac{R}{R_{e,X}}\right)^{1/n_X}  
\label{eq:sersic}
\end{equation}
where $\mu_{0,X}$ is the central surface brightness of the S\'ersic
model in the passband $X$; $n_X$ is the corresponding S\'ersic (shape)
parameter; and $b_{n,X} \sim 2 n_X -1/3$ is a constant defined so that
$R_{e,X}$ is the half-light radius of the model~\citep{CCD93}. The
radius, $R$, is the circularized galactocentric distance .  For each
galaxy we construct the colour profile, in the form $g-X$ (with
$X=rizYJHK$), by computing the difference of the corresponding S\'ersic
profiles:
\begin{eqnarray}
g \! - \! X \, (R)  =  \mu_{0,g} - \mu_{0,X} +  \nonumber 
\end{eqnarray}
\begin{equation}
+  1.0857 \cdot \left[ b_{n,g} \left(\frac{R}{R_{e,g}}\right)^{1/n_g} - b_{n,X} \left(\frac{R}{R_{e,X}}\right)^{1/n_X}  
\right]. 
\label{eq:col_sersic}
\end{equation}
All the parameters $\mu_{0,X}$, $R_{e,X}$, and $n_X$, are obtained by
fitting the galaxy image in waveband $X$ with a PSF-convolved S\'ersic
model, as detailed in Paper I.  For a given galaxy mass bin, a stacked
colour profile is then computed. To this effect, for a given galaxy mass 
bin, the galactocentric radii of each individual
colour profile are normalized by the corresponding $R_{e,r}$, and the median of
all the so-normalized colour profiles is computed.


\subsection{Systematics}
\label{sec:systematics}
At the very faint surface brightness levels we want to explore
(i.e. $\simgt 25\,\rm mag\,arcsec^{-2}$ in the $r$ band, at distances
larger than a few effective radii, see Fig.~\ref{fig:profs}), the
colour profile of ETGs can be affected by several sources of
systematics. We discuss here the role of such systematics, and how we
take them into account in the analysis.

{\it Background subtraction.} As described in Paper I, { structural
  parameters of ETGs  are estimated by fitting galaxy  images } with a
PSF-convolved S\'ersic  model plus  a constant value  representing the
background.  An  error in  the background estimate  can bias  the $\rm
R_{e,X}$  and  $n_X$ parameters,  changing  the  shape  of the  colour
profiles (see  Eq.~\ref{eq:col_sersic}). To quantify  the relevance of
this  effect, we have  repeated the  S\'ersic fitting  in $g$  and $r$
bands, fixing  the background to  the value measured in  the outermost
regions  of  each  postage   stamp  frame  by  applying  the  biweight
statistics~\citep{Beers:90}.  We  found that the  stacked $g-r$ colour
profile does  not change  significantly when using  a fixed-background
fitting { (with respect to  the case where background is treated as
  a free fitting  parameter)}, with a variation $< 0.02  \rm \, mag \,
arcsec^{-2}$ at the largest radii probed ($\rm R/R_e=8$), i.e. smaller
than the  typical statistical error  of the stacked profiles  at these
radii   ($\simgt    0.03   \rm    \,   mag   \,    arcsec^{-2}$;   see
Tables~\ref{tab:colours_lowmass},           ~\ref{tab:colours_intmass},
and~\ref{tab:colours_highmass}).  { In this work, we use structural
  parameters,  and  hence  color  profiles, derived  by  treating  the
  background  as a  free  fitting parameter  (i.e.  adopting the  same
  procedure as in Paper I) }.

\begin{figure}
\begin{center}
\includegraphics[height=80mm]{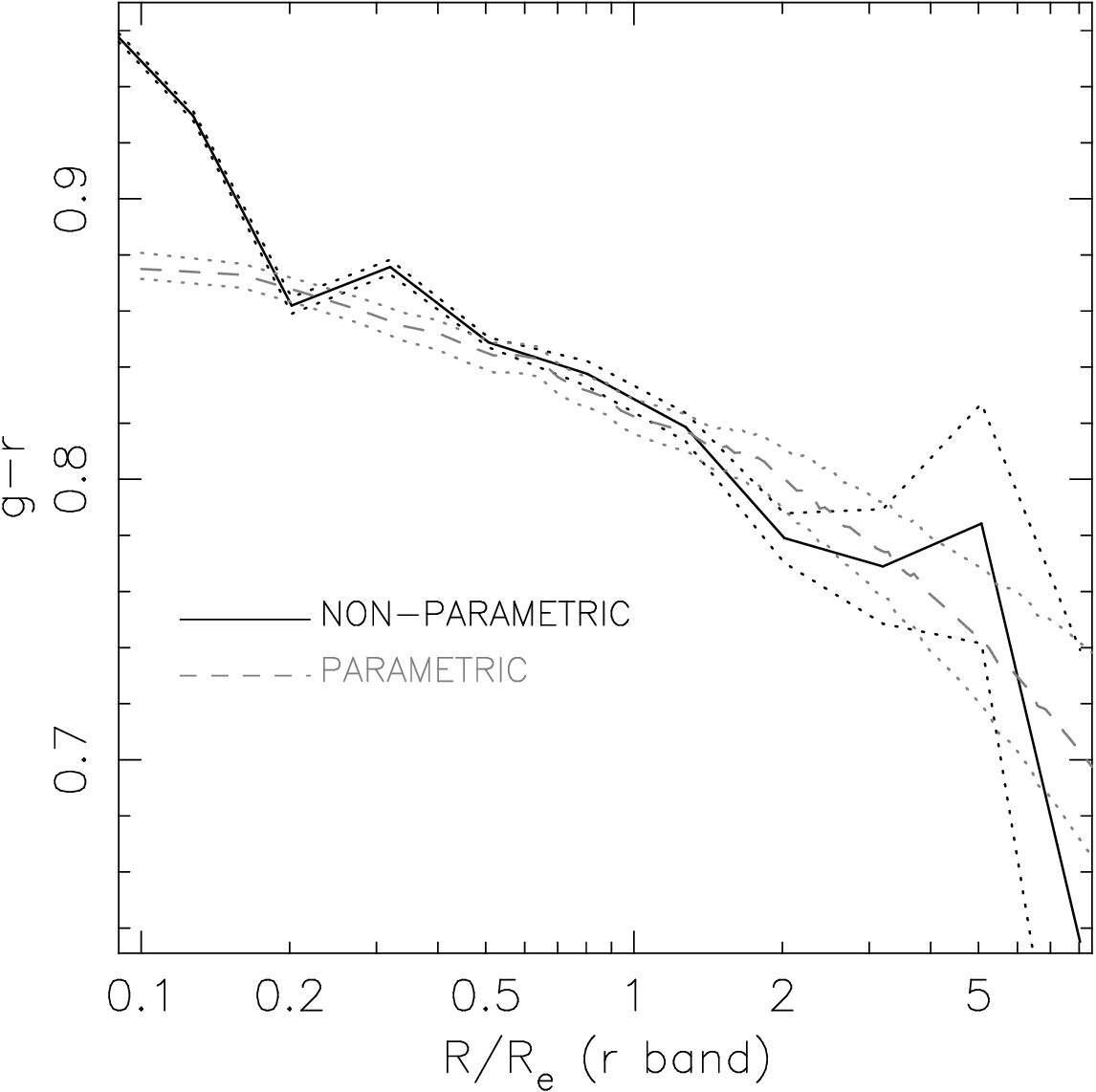}
\end{center}
\caption{{ Median colour profiles  of high-mass ETGs as obtained by
    the  parametric (dashed  curve) and  non-parametric  (solid curve)
    approaches (see text for details). Dotted grey (black) curves mark
    the  $\pm  1$~$\sigma$ error  on  the parametric  (non-parametric)
    profile.   The bump  in the  inner ($\rm  R/R_e<0.2$) part  of the
    non-parametric profile is due to  the variation of the PSF between
    the  $g$  and $r$  bands  (the latter  having  a  a smaller  FWHM,
    producing  the red  bump).   The profiles  are  in good  agreement
    within the scatter out to  the largest radii probed in the present
    study  ($\sim  8\,\rm  R_e$).   Notice  how  the  scatter  in  the
    non-parametric   profile    increases   significantly   at   large
    galactocentric distances.}  }
\label{fig:par_nopar}
\end{figure}

{\it  Parametric vs.   non-parametric profiles.}   Our  stacked colour
profiles are obtained by a parametric approach, using the best-fitting
S\'ersic  models   in  different  wavebands  to   compute  the  colour
profiles. {  This approach is well  justified by the  fact that, on
  average,  the  light  profiles   of  massive  ellipticals  are  well
  described by a single S\'ersic  model out to roughly eight effective
  radii~\citep{Kormendy:09, TalvanDokkum:11}.   Moreover, the $\chi^2$
  selection (see Sec.~\ref{sec:samples}) allows us to exclude galaxies
  whose light profile is not  reproduced by the S\'ersic model.  These
  objects  are  mostly  early-spiral  contaminants (see  Paper  I  for
  details).  In App.~\ref{sec:light_profiles} we  show the  quality of
  our  S\'ersic fits,  by comparing  the (stacked)  best-fitting light
  profiles,  derived  on   circular  apertures,  with  those  measured
  directly from the galaxy images, for all relevant wavebands and each
  mass bin.  Parametric and  non-parametric light profiles turn out to
  be   fairly  consistent,   within   uncertainties,  supporting   the
  robustness  of   our  measurements  out   to  large  galacto-centric
  distances.  A further test of our parametric approach is illustrated
  in  Fig.~\ref{fig:par_nopar},  comparing,  as example,  the  stacked
  $g-r$  profiles  obtained   by  the  parametric  and  non-parametric
  approaches  for  high-mass  ETGs.   Our  choice  of  $g-r$  for  the
  comparison, rather  than other colour profiles, is  justified by the
  fact that this colour is less sensitive to the effect from the wings
  of the  PSF (see  below).  Also, because  of the  shorter wavelength
  baseline  with  respect  to  other  colors, $g-r$  provides  a  more
  critical test.  For each galaxy, we measured  the surface brightness
  profiles in the $g$ and $r$  bands directly from the images over the
  same concentric ellipses in both bands, all ellipses having the same
  axis  ratio from the  $r$-band S\'ersic  fit and  circularized radii
  equally spaced by $0.5$pixel (i.e.  $\sim 0.03 \, R_e$ for high-mass
  ETGs).  For each galaxy, we  then rescaled the radii of the ellipses
  by a factor $1/R_e$, where  $R_{e,r}$ is the r-band effective radius
  of the given galaxy.   The rescaled profiles were median-combined in
  both $g$ and $r$ bands, to generate the non-parametric $g-r$ stacked
  color profile. The stacked  non-parametric profile was then averaged
  over radial  bins logarithmically spaced by  $0.15$~dex, the profile
  uncertainty in  each bin being computed from  the standard deviation
  of colour  values in  the bin.  Fig.~\ref{fig:par_nopar}  shows that
  parametric and  non-parametric profiles differ  significantly } only
in  the  inner  part,  at  $\rm R/R_e<0.2$,  as  expected,  since  the
non-parametric profile is not corrected for seeing effects, while they
agree quite  well out to  a galactocentric distance of  $\sim 3-4\,\rm
R_e$.  We decide  to continue our study using  the parametric approach
(i.e.   S\'ersic fitting)  as the  corresponding uncertainties  in the
surface brightness photometry are  significantly smaller at large $\rm
R/R_e$, with  respect to the  case where a non-parametric  approach is
adopted -- a crucial aspect for the present study.

\begin{figure}
\begin{center}
\includegraphics[height=120mm]{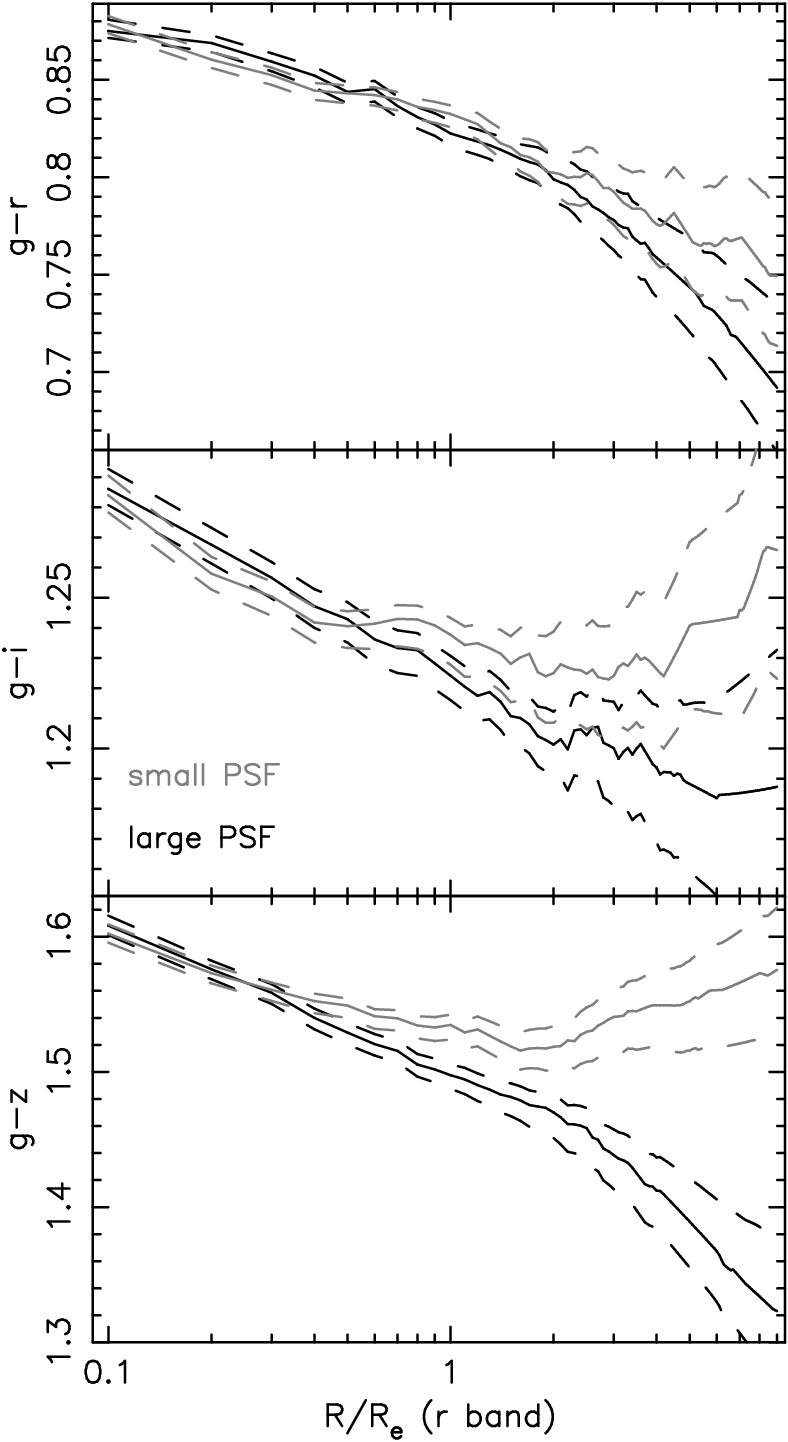}
\end{center}
\caption{Comparison of $g-r$, $g-i$, and $g-z$, colour profiles, when
  using a small ($r_{\rm PSF}=2.8^{\prime\prime}$, grey curves) and large
  ($r_{\rm PSF}=24^{\prime\prime}$, black curves) PSF model (see text for
  details). Solid curves mark median trends, while dashed curves are
  the $\pm 1\sigma$ uncertainties on the median.}
\label{fig:small_large_PSF}
\end{figure}

{\it PSF red halos.} The shape of the PSF at large radii can alter the
colour profile of a galaxy. As regards SDSS data, the PSF exhibits
prominent wings at distances $\simgt 10^{\prime\prime}$, being more
pronounced in the $i$ band than in the $g$ and $r$
filters~\citep{BZC:10}. If neglected, the wings in the redder bands of
the PSF cause a spurious red colour excess (in $g-i$ and $r-i$) in the
galaxy outskirts~\citep{deJong:08, TalvanDokkum:11}. Our colour
profiles are based on PSF-corrected structural parameters. The PSF is
obtained by fitting images of stars with a superposition of 2D Moffat
functions, truncated at a maximum radius, $r_{\rm PSF}$, where the PSF
model reaches a mean flux level of $1\%$ of its central value. In the
previous papers of the SPIDER survey, $r_{\rm PSF}$ is typically $\sim
7$ pixels (i.e. $\sim 2.8^{\prime\prime}$) in all wavebands. This
issue is not relevant for those papers, as only the properties within
1--2 effective radii were considered.  However, when studying the
profiles out to $8\,\rm R_e$, this choice will underestimate the
effect of the wings of the PSF. To account for this, we have
reprocessed the $griz$ images of all the $674$ selected ETGs using a
large PSF model, with $r_{\rm PSF}=60$~pixels ($\sim
24^{\prime\prime}$).  Hereafter, we refer to the original ($r_{\rm
  PSF}=7$~pixels) and the updated PSF models as the small and large
PSFs, respectively. Fig.~\ref{fig:small_large_PSF} compares the
median-stacked $g-r$, $g-i$, and $g-z$, colour profiles of high-mass
ETGs with respect to the size of the PSF model. Using a small PSF
makes the galaxy colour redder in the external regions, beyond $\sim
1\,\rm R_e$, in agreement with \citet{deJong:08} and
\citet{TalvanDokkum:11}. As expected from the waveband dependence of
the PSF halo (see fig.~6 of~\citealt{BZC:10}), the effect is stronger
in $g-i$ (and $g-z$) than $g-r$.  In the inner regions, at $\rm
R/R_e<1$, the profiles (and hence the inner colour gradients) are not
affected by the choice of PSF model. For the analysis presented in
this paper, we have obtained all optical colour profiles using the large PSF.
Notice that UKIDSS PSFs are not affected from any
extended wing, as oppose to the i- and z-band (SDSS) PSFs. Also, the large
wavelength baseline probed by optical--NIR colors makes them 
insensitive to any PSF red halo effect. For this reason, we
have decided not to reprocess the NIR wavebands, and use the 
Sersic profiles from our previous works.  

\begin{figure*}
\begin{center}
\includegraphics[height=80mm]{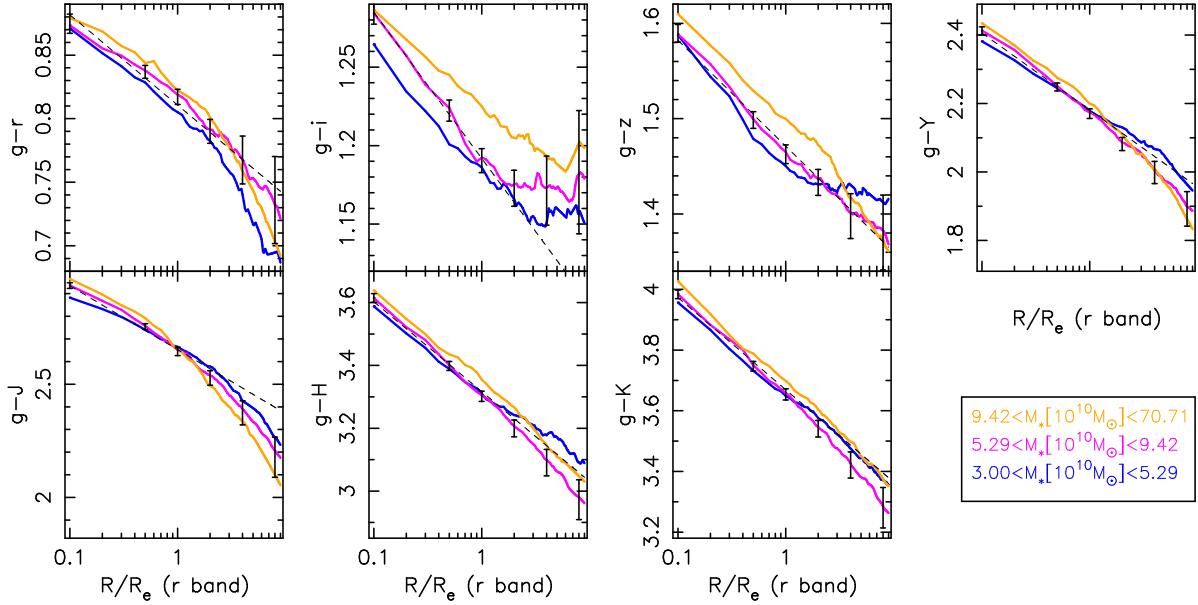}
\end{center}
\caption{Median colour profiles, in the form of $g-X$ (with
  $X=rizYJHK$), for the sample of $674$ ETGs, as a function of the
  normalized galactocentric distance $\rm R/R_e$. For each panel, the
  blue, magenta, and orange curves are the median profiles for
  galaxies in the same stellar mass bins, as labelled (see bottom--right of the Figure). The black
  dashed lines have slopes equal to the peak values of colour gradient
  distributions of SPIDER ETGs, as reported in tab.~2 of Paper IV. The
  intercept of the dashed lines has been arbitrarily chosen to match
  the inner ($\rm R<R_e$) colour profile of intermediate mass galaxies
  (magenta curve). The six vertical error bars mark the positions of
  $0.1, 0.5, 1, 2, 4, 8$ effective radii that we select to analyze the
  variation of colour gradients in ETGs from their centre to the
  outskirts. The size of the error bars is given by the $\pm 1 \sigma$
  error on the median colour profile of intermediate mass galaxies
  (magenta colour).  Notice the remarkable upturn in the $g-i$ and
  $g-z$ profiles at $\rm R\simgt R_e$ for intermediate- and
  high-mass galaxies.  } ~\label{fig:col_profs}
\end{figure*} 

\section{Colour profiles vs. galaxy mass}
\label{sec:col_prof_mass}
Fig.~\ref{fig:col_profs} plots the median colour profiles, $g-X$ (with
$X=rizYJHK$), obtained  by median-stacking the single  profiles of all
galaxies in each of the three  stellar mass bins.  For each panel, the
black dashed  line represents the slope  of the colour  profile in the
radial  range from  $0.1$ to  $1\,\rm R_e$,  i.e.  the  average colour
gradient,  $\nabla_{g-X}$, as estimated  in Paper  IV. Error  bars are
$1\sigma$ uncertainties on median colours, for intermediate-mass ETGs,
at different fiducial galactocentric distances, i.e. $\rm R/R_e=\{0.1,
0.5, 1, 2,  4, 8\}$.  The values of $0.1$ and  $1\,\rm R_e$ define the
range usually  adopted to estimate  internal colour gradients  of ETGs
(e.g.~\citealt{PVJ90}), while the distances $\{2, 4, 8\}\,\rm R_e$ are
chosen to analyze  the behaviour of colour gradients  in the outermost
regions.   We adopt  a maximum  radius of  $8\,\rm R_e$,  so  that the
photometric errors are kept  small enough ($\simlt 0.04$~mag) to allow
for      a       meaningful      stellar      population      analysis
(Sec.~\ref{subsec:color_fitting}).   Moreover,  as  noticed  above,  a
single S\'ersic model might not  be able to describe the light content
of  ETGs at  larger  galactocentric distances~\citep{TalvanDokkum:11}.
The median $g-X$ colours  (with $X=rizYJHK$), computed at the fiducial
galactocentric   distances,   are   reported,  together   with   their
uncertainties,                                                       in
Tables~\ref{tab:colours_lowmass},~\ref{tab:colours_intmass},
and~\ref{tab:colours_highmass}, for low-, intermediate-, and high-mass
ETGs, respectively.   { Notice that  both the median  $g-X$ colours
  and  colour uncertainties  are obtained  by stacking  the parametric
  color profiles (Eq.~\ref{eq:col_sersic}). }

\begin{figure}
\begin{center}
\includegraphics[height=75mm]{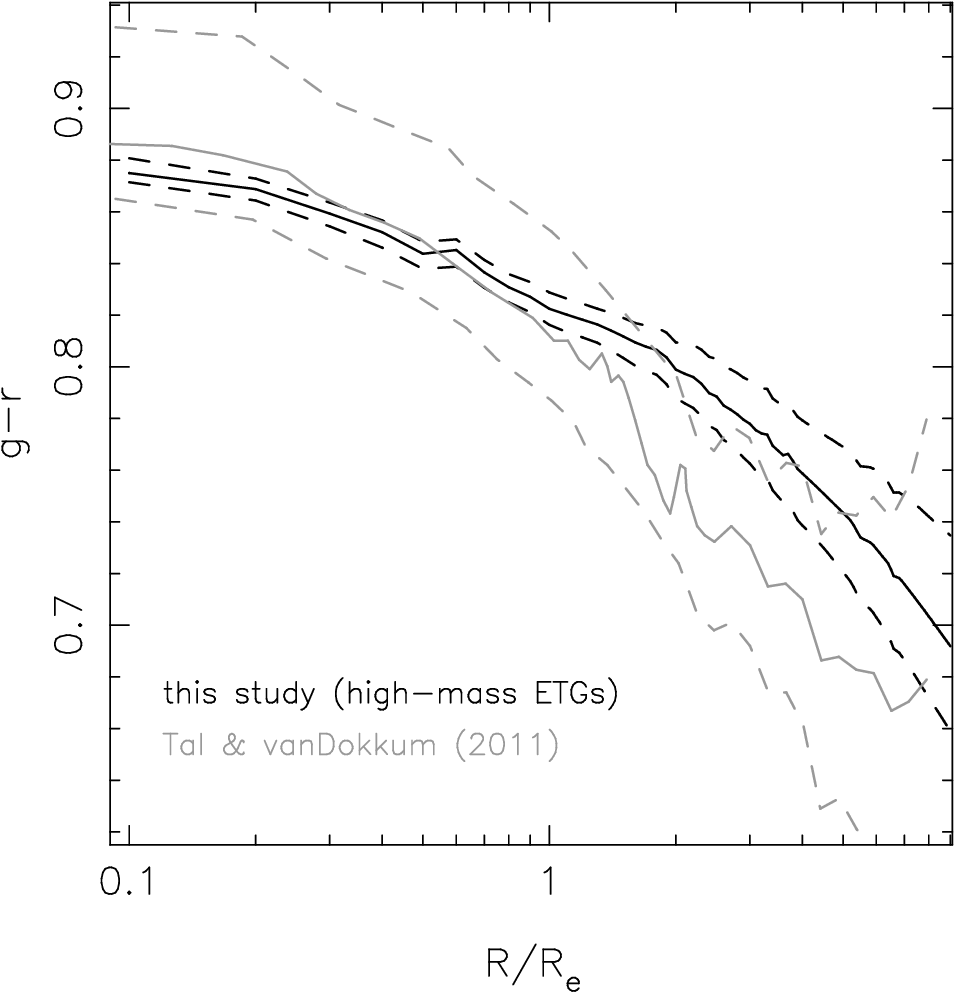}
\end{center}
\caption{Comparison of our $g-r$ stacked colour profile with the $r-i$
  stacked profile obtained for ETGs at redshift z$\sim 0.34$ by
  TvD11. Notice that at z$\sim 0.34$ the $r-i$ colour corresponds
  approximately to $g-r$ in the rest frame. The TvD11 profile matches well our
  colour profile in the inner galaxy region ($\rm R < R_e$), while at
  larger radii their profile appears bluer (by $\sim 0.03$~mag), although
  fully consistent within the error bars, with our results.}
\label{fig:comp_TvD11}
\end{figure} 

Both  optical--optical and optical--NIR  colour profiles  become bluer
towards the galaxy outskirts, up  to $\sim 8\,\rm R_e$.  Overall, with
the exception of  $g-r$ and $g-J$, the profiles  are reasonably linear
over the entire log-radial range. In $g-i$, low- and intermediate-mass
galaxies  exhibit  some  deviation  from  the  linear  trend  at  $\rm
R/R_e\simgt 3$,  perhaps reflecting some inaccuracy in  the i-band PSF
halo modeling at large distances. However, the deviation is comparable
to the size  of the error bars, making the  effect unimportant for the
stellar  population  analysis (Sec.~\ref{subsec:color_fitting}).   The
curvature  in the  $g-r$ profiles  is small,  considering  error bars,
while the  $g-J$ profile shows a  strong curvature, not  seen in other
optical--NIR  colours  ($g-Y$,  $g-H$,   and  $g-K$).  This  might  be
explained by  systematics in the J-band structural  parameters, due to
the  different quality and  resolution of  the J-band  photometry (see
Paper I for  more details). { For this reason, we  do not use $g-J$
  to     constrain radial trends of    stellar     properties
  (Sec.~\ref{subsec:color_fitting}) }.

This paper constitutes the first comprehensive analysis of
optical--optical and optical--NIR colours, as a function of galaxy
mass, out to the halo region of ETGs ($8\,\rm R_e$). On the other hand, for
what concerns optical data alone, we can compare our results with
those obtained by~\citet[hereafter TvD11]{TalvanDokkum:11}, whose
analysis was based on stacking a large sample of $42,000$ luminous red
galaxies from SDSS, at redshift z$\sim 0.34$.
Fig.~\ref{fig:comp_TvD11} compares our $g-r$ colour profile, for
high-mass ETGs, with their $r-i$ stacked profile.  Notice that, at
z$\sim 0.34$, the $r-i$ colour roughly corresponds to $g-r$ in the
rest-frame, allowing for a direct comparison to our colour
profiles. The $r-i$ profile has been extracted from fig.~8 of TvD11
(left panel) and normalized to match our average $g-r$ within $\rm
R_e$. Our profile and that of TvD11 are in very good agreement within
an effective radius, while at large distances, the TvD11 stacked data
appear slightly bluer (by $\sim 0.03$~mag), but fully consistent
with our results, considering error bars and the different selection criteria (e.g.. 
redshift range) of our and TdV11 samples.

\begin{table*}
\centering
\small
\begin{minipage}{220mm}
\caption{Median colours of low-mass ETGs at different galactocentric
  distances.}
\begin{tabular}{c|c|c|c|c|c|c|c}
\hline
 $\rm R/R_e$ & $g-r$ & $g-i$ & $g-z$ & $g-Y$ & $g-J$ & $g-H$ & $g-K$ \\
\hline 
$  0.1$ & $ 0.871 \pm 0.009$ & $ 1.264 \pm 0.011$ &$ 1.588 \pm 0.011$ &$ 2.382 \pm 0.015$ &$ 2.882 \pm 0.015$ &$ 3.588 \pm 0.016$ &$ 3.957 \pm 0.017$  \\  
$  0.5$ & $ 0.829 \pm 0.005$ & $ 1.201 \pm 0.006$ &$ 1.479 \pm 0.008$ &$ 2.245 \pm 0.010$ &$ 2.741 \pm 0.013$ &$ 3.389 \pm 0.012$ &$ 3.732 \pm 0.014$  \\  
$  1.0$ & $ 0.805 \pm 0.006$ & $ 1.186 \pm 0.007$ &$ 1.450 \pm 0.009$ &$ 2.178 \pm 0.012$ &$ 2.665 \pm 0.015$ &$ 3.311 \pm 0.014$ &$ 3.649 \pm 0.016$  \\  
$  2.0$ & $ 0.782 \pm 0.008$ & $ 1.166 \pm 0.009$ &$ 1.431 \pm 0.012$ &$ 2.130 \pm 0.016$ &$ 2.579 \pm 0.026$ &$ 3.242 \pm 0.019$ &$ 3.576 \pm 0.026$  \\  
$  4.0$ & $ 0.740 \pm 0.015$ & $ 1.151 \pm 0.017$ &$ 1.431 \pm 0.022$ &$ 2.067 \pm 0.029$ &$ 2.427 \pm 0.043$ &$ 3.178 \pm 0.031$ &$ 3.475 \pm 0.040$  \\  
$  8.0$ & $ 0.695 \pm 0.028$ & $ 1.157 \pm 0.033$ &$ 1.412 \pm 0.040$ &$ 1.964 \pm 0.048$ &$ 2.263 \pm 0.072$ &$ 3.091 \pm 0.054$ &$ 3.372 \pm 0.062$  \\  
\hline
\end{tabular}
\label{tab:colours_lowmass}
\end{minipage}
\end{table*}

\begin{table*}
\centering
\begin{minipage}{220mm}
\small
\caption{Median colours of intermediate-mass ETGs at different
  galactocentric distances.}
\begin{tabular}{c|c|c|c|c|c|c|c}
\hline
 $\rm R/R_e$ & $g-r$ & $g-i$ & $g-z$ & $g-Y$ & $g-J$ & $g-H$ & $g-K$ \\
\hline 
$  0.1$ & $ 0.873 \pm 0.008$ & $ 1.284 \pm 0.009$ &$ 1.588 \pm 0.009$ &$ 2.413 \pm 0.011$ &$ 2.936 \pm 0.012$ &$ 3.615 \pm 0.014$ &$ 3.984 \pm 0.014$  \\  
$  0.5$ & $ 0.838 \pm 0.005$ & $ 1.224 \pm 0.007$ &$ 1.500 \pm 0.009$ &$ 2.251 \pm 0.011$ &$ 2.758 \pm 0.013$ &$ 3.401 \pm 0.014$ &$ 3.750 \pm 0.016$  \\  
$  1.0$ & $ 0.819 \pm 0.006$ & $ 1.192 \pm 0.008$ &$ 1.465 \pm 0.010$ &$ 2.174 \pm 0.014$ &$ 2.651 \pm 0.019$ &$ 3.305 \pm 0.016$ &$ 3.659 \pm 0.018$  \\  
$  2.0$ & $ 0.792 \pm 0.009$ & $ 1.175 \pm 0.012$ &$ 1.437 \pm 0.014$ &$ 2.086 \pm 0.019$ &$ 2.539 \pm 0.032$ &$ 3.207 \pm 0.027$ &$ 3.546 \pm 0.027$  \\  
$  4.0$ & $ 0.768 \pm 0.019$ & $ 1.173 \pm 0.022$ &$ 1.401 \pm 0.024$ &$ 2.006 \pm 0.032$ &$ 2.391 \pm 0.053$ &$ 3.099 \pm 0.042$ &$ 3.426 \pm 0.043$  \\  
$  8.0$ & $ 0.732 \pm 0.034$ & $ 1.182 \pm 0.039$ &$ 1.383 \pm 0.039$ &$ 1.899 \pm 0.050$ &$ 2.206 \pm 0.089$ &$ 2.981 \pm 0.063$ &$ 3.285 \pm 0.067$  \\  
\hline
\end{tabular}
\label{tab:colours_intmass}
\end{minipage}
\end{table*}

\begin{table*}
\centering
\begin{minipage}{220mm}
\small
\caption{Median colours of high-mass ETGs at different galactocentric
  distances.}
\begin{tabular}{c|c|c|c|c|c|c|c}
\hline
 $\rm R/R_e$ & $g-r$ & $g-i$ & $g-z$ & $g-Y$ & $g-J$ & $g-H$ & $g-K$ \\
\hline 
$  0.1$ & $ 0.879 \pm 0.005$ & $ 1.286 \pm 0.006$ &$ 1.610 \pm 0.007$ &$ 2.434 \pm 0.010$ &$ 2.965 \pm 0.012$ &$ 3.639 \pm 0.013$ &$ 4.027 \pm 0.014$  \\  
$  0.5$ & $ 0.844 \pm 0.005$ & $ 1.245 \pm 0.006$ &$ 1.529 \pm 0.008$ &$ 2.275 \pm 0.011$ &$ 2.792 \pm 0.014$ &$ 3.435 \pm 0.013$ &$ 3.789 \pm 0.015$  \\  
$  1.0$ & $ 0.822 \pm 0.006$ & $ 1.225 \pm 0.007$ &$ 1.500 \pm 0.009$ &$ 2.200 \pm 0.014$ &$ 2.650 \pm 0.020$ &$ 3.355 \pm 0.017$ &$ 3.699 \pm 0.020$  \\  
$  2.0$ & $ 0.801 \pm 0.010$ & $ 1.205 \pm 0.011$ &$ 1.473 \pm 0.015$ &$ 2.110 \pm 0.023$ &$ 2.492 \pm 0.034$ &$ 3.267 \pm 0.027$ &$ 3.604 \pm 0.029$  \\  
$  4.0$ & $ 0.759 \pm 0.020$ & $ 1.193 \pm 0.022$ &$ 1.416 \pm 0.026$ &$ 2.009 \pm 0.041$ &$ 2.336 \pm 0.063$ &$ 3.145 \pm 0.045$ &$ 3.498 \pm 0.048$  \\  
$  8.0$ & $ 0.703 \pm 0.035$ & $ 1.202 \pm 0.040$ &$ 1.373 \pm 0.046$ &$ 1.863 \pm 0.064$ &$ 2.099 \pm 0.099$ &$ 3.047 \pm 0.072$ &$ 3.375 \pm 0.073$  \\  
\hline
\end{tabular}
\label{tab:colours_highmass}
\end{minipage}
\end{table*}

\begin{figure}
\begin{center}
\includegraphics[height=75mm]{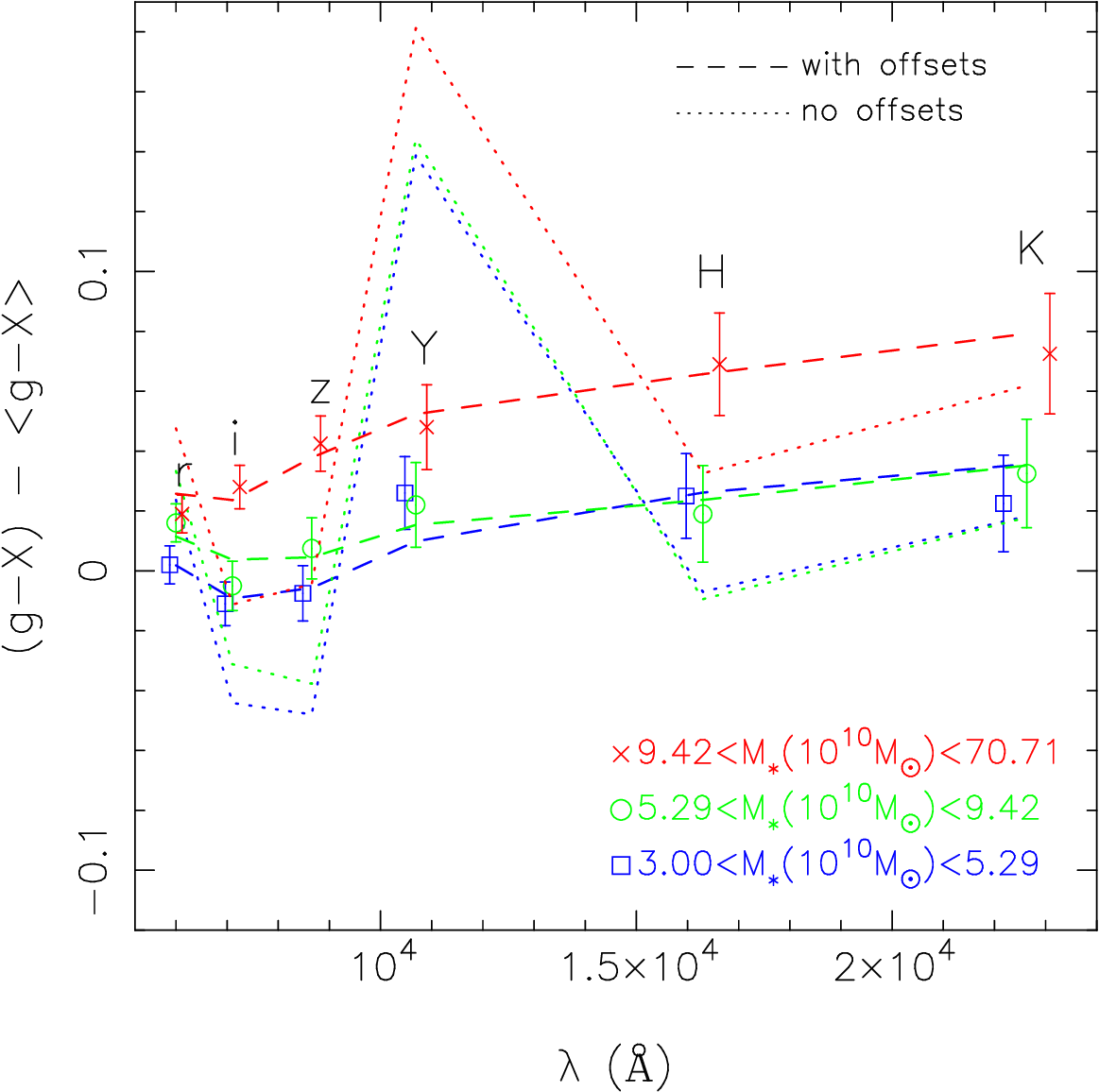}
\end{center}
\caption{The median colours of ETGs, computed at $\rm R=R_e$, are plotted
  in the form of $(g-X)-\langle g-X\rangle$ (with $ X=rizYHK$) as  a function of the
  typical wavelength of filter $X$, where $\langle g-X\rangle$ is the median of all
  $g-X$ values at different galaxy radii and masses
  (Tables~\ref{tab:colours_lowmass}, ~\ref{tab:colours_intmass},
  and~\ref{tab:colours_highmass}). Different symbols refer to
  galaxies with different mass, as indicated in the lower-right of the
  plot. Error bars are $1\sigma$ uncertainties on median colours.
  Dotted curves are the best-fitting SSP models from CB$^{\ast \!}$MC (see 
  text). Notice the discrepancy between models and data.
  The dashed curves are best-fitting SSP models after suitable
  correction offsets ($OFF_X$) were applied, improving
  the matching between models and data.}
\label{fig:col_profs_offsets}
\end{figure} 

\begin{figure}
\begin{center}
\includegraphics[height=120mm]{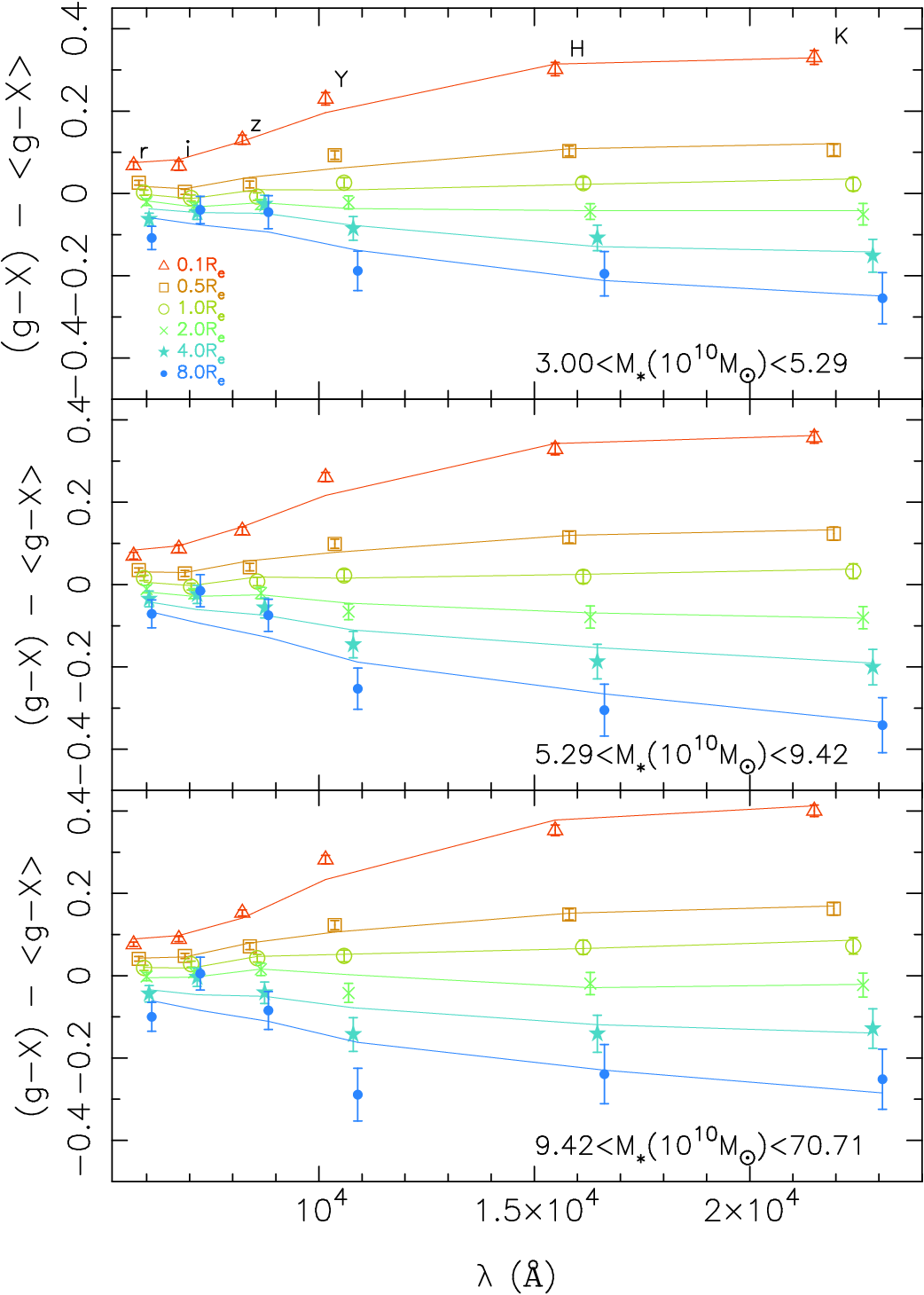}
\end{center}
\caption{The colours used in this paper ($g-X$) are shown as a
  function of the effective wavelength of filter $X$. Top, middle, and
  bottom panels correspond to low-, intermediate-, and high-mass
  galaxies, respectively. For each mass bin, different
  galactocentric distances, $\rm R/R_e=\{0.1, 0.5, 1, 2, 4, 8\}$, are
  plotted with different colours and symbols (see lower-left of top
  panel).  The curves are best-fitting SSP colours for CB$^\ast \!$MC models.
  Notice the good matching of observed and model colours.  }
\label{fig:col_profs_fits}
\end{figure}

\section{Constraining the properties of the underlying stellar populations}
\label{sec:fitting}

\subsection{Stellar population models}
\label{subsec:SPmodels}
We compare the observed colours of ETGs with the predictions of three different
versions of a well established stellar population (SP) synthesis code, 
{\it (a)} the \citet[][hereafter BC03]{BrC03} models,
{\it (b)} the minor revision of these models introduced by \citet[][henceforth CB07]{CBr07}, and
{\it (c)} the major revision of this code and models in preparation by \citet[][hereafter CB$^\ast$]{CBr13}.
The CB07 models use the same sets of stellar tracks and spectral libraries as the BC03
models (see BC03 for details), except for the thermally pulsing asymptotic giant branch (TP-AGB) stars. 
For these stars, CB07 follow the semi-empirical evolutionary prescriptions by \citet[]{MG07} and \citet[]{MG08}.
The CB$^\ast$ population synthesis models used in this paper are based on the stellar 
evolution models computed with updated input physics for stars up to $15 M_\odot$ by \cite{bertl08}.
These tracks are available for metallicities 
$Z =\{0.0001, 0.0004, 0.001, 0.002, 0.004, 0.008, 0.017, 0.04, 0.07\}$, with the solar 
metallicity being $Z_\odot=0.017$.  
In CB$^\ast$ the evolution of TP-AGB stars follows a recent prescription by 
Marigo \& Girardi (private communication),
which has been calibrated using observations of AGB stars in the Magellanic Clouds and nearby 
galaxies~\citep{gir10, melb12}. In the optical range, the CB$^\ast$ models are available for the 
IndoUS \citep{fv04}, Miles \citep{ps06}, Stelib \citep{jfl03}, and BaSeL 3.1\citep{pw02} spectral libraries.
The NIR spectra of TP-AGB stars in CB$^\ast$ are selected from the compilation by \cite{lm02}, 
the IRTF library \citep{jr09}, and the C-star model atlas by \cite{ba09}.
For stars of metallicity near $Z_\odot$, the number of available stellar spectra in
the IndoUS and Miles libraries increases by a factor of roughly 20 with respect to the Stelib library used in BC03.
Thus, both in metallicity and spectral coverage, the CB$^\ast$ models represent a major improvement
over the BC03 models, which translate into a better modelling of the NIR and optical galaxy colours.

Recently, it has  become clear that the treatment  of the TP-AGB stars
in  the CB07  models tend  to overestimate  the contribution  by these
stars in the NIR \citep{kriek10,melb12,zib12}.  We use the CB07 models
in  this  paper for  the  sake  of  completeness and  comparison  with
previous work (e.g.~\citealt{LdC:09}, and Paper IV).
   
The  simple stellar  population (SSP)  models adopted  in  the present
study,  and the  labels we  use to  refer to  them, are  summarized in
Tab.~\ref{tab:models}.   The  IMFs  used  are Scalo  (BC03),  Chabrier
(CB07), while we consider  both Chabrier and Salpeter~\footnote{i.e. a
  power-law IMF with slope $1.35$, and lower and upper mass cutoffs of
  $0.1$   and  $100  \,   M_\odot$,  respectively.}~(1955)   IMF,  for
CB$^\ast$.  Models  in the metallicity  range between $1/50$  and $2.5
Z_{\odot}$ are considered, in the age range from $1$ to $18$~Gyr. {
  The lower age  limit of $1$~Gyr is because  we study ETGs, dominated
  by old  stellar population, while the  upper limit of  18~Gyr is the
  maximum age  value available in  the models~\footnote{ We  do not
    set the  age of  the Universe  as an upper  limit for  the present
    analysis, as  the absolute matching  of model and  observations is
    still  very uncertain.  In fact,  large differences  exist between
    best-fitting      ages     from     different      models     (see
    Sec.~\ref{subsec:SSP_gradients}),  while  relative differences  in
    age and  metallicity (i.e.  the radial gradients)  are much better
    constrained.}.}

For a given model, synthetic colours are computed for three different star formation histories:
\begin{description}
 \item[- {\it Simple Stellar Population.}] Equivalent to an instantaneous burst of star formation. 
 \item[- {\it $\tau$ models.}] We assume an exponentially declining
   star formation rate with e-folding time $\tau$.  The age of the
   model (i.e. the first epoch of star formation) and its metallicity
   are in the same ranges as for the SSPs. The $\tau$ is varied from
   zero to the age of the model.
 \item[- {\it Burst model.}] The star formation rate is constant, with
   duration $\Delta t$. The age and metallicity of the model are
   varied in the same ranges as for $\tau$ models. $\Delta t$ is
   varied from zero to the age of the model.
\end{description}
To compute synthetic colours, $(g-X)_{MOD}$\ ($ X=rizYHK$, see below),
each SP model is converted to the median redshift of ETGs in our
sample ($z=0.0614$), and integrated with the $ grizYHK$ throughput
curves (see Paper I for details).

\begin{table*}
\centering
\small
\caption{Stellar population synthesis models adopted in the present study.}
\begin{tabular}{lll}
\hline
model & stellar library & IMF \\
\hline
BC03                       & STELIB    & Scalo \\
CB07SC                  & STELIB    & Chabrier \\
CB$^\ast \!$IC(CB$^\ast \!$IS)     & INDOUX  & Chabrier (Salpeter) \\
CB$^\ast \!$MC(CB$^\ast \!$MS) & MILES     & Chabrier (Salpeter) \\
CB$^\ast \!$SC(CB$^\ast \!$SS)  & STELIB    & Chabrier (Salpeter) \\
CB$^\ast \!$BC(CB$^\ast \!$BS)  & BaSel 3.1 & Chabrier (Salpeter) \\
\hline
\end{tabular}
\label{tab:models}
\end{table*}


\subsection{Colour fitting}
\label{subsec:color_fitting}

For a given galaxy mass bin, and given galactocentric distance, we fit the 
corresponding median colours of  ETGs by minimizing the following expression:
\begin{equation}
\chi^2= \sum_X \frac{ \left[ (g-X)_{OBS}-(g-X)_{MOD}+OFF_X \right]^2 }{\sigma_{g-X}^2},
\label{eq:chi}
\end{equation}
where  the   summation  extends  over  the   available  filters  ($
X=rizYHK$)  , and  the subscripts  $OBS$ and  $MOD$ refer  to observed
(Tables~\ref{tab:colours_lowmass},          ~\ref{tab:colours_intmass},
and~\ref{tab:colours_highmass})  and  synthetic  (model) colours  (see
Sec.~\ref{subsec:SPmodels}),  respectively.  The  terms  $OFF_X$ ($
X=rizYHK$) are photometric zero-point  offsets, adopted to improve the
data  and model matching  (see below).  In practice,  the minimization
consists  of searching for  the set  of stellar  population parameters
(e.g.  the   age  and  metallicity  for  SSP   models)  that  minimize
Eq.~\ref{eq:chi}. We remind the reader that $g-J$ colours are not used
in  the   minimization  procedure,   for  the  reasons   explained  in
Sec.~\ref{sec:col_prof_mass}.   Because  some  model   parameters  are
intrinsically  degenerate when  fitting broad-band  colours  (e.g. the
formation epoch  and $\tau$,  or $\Delta t$),  for each model  we give
only the mass-weighted age parameter ($Age$, hereafter simply referred
to as  the age of  the model), and  metallicity ($Z$) and  analyze how
these   quantities  change   with  galaxy   mass   and  galactocentric
distance. This allows  us to verify how the assumption  of a given SFR
affects the estimate of age and metallicity profiles in ETGs.

The terms  $OFF_X$ in Eq.~\ref{eq:chi}  depend only on the  adopted SP
synthesis  code (although  the dependence  is small,  see  below), and
account  for the fact  that, in  general, model  colours do  not match
exactly the  observed colours, because of, e.g.,  uncertainties in the
photometric   calibration  and/or   SP   synthesis  ingredients.    To
illustrate this point,  we consider CB$^\ast \!$MC models  and plot in
Fig.~\ref{fig:col_profs_offsets}, the $\rm  R=1\,R_e$ colours of low-,
intermediate-, and high-mass ETGs.  For a clearer figure including all
colours  in the  same plot,  we  have subtracted  the median  colours,
$\langle g-X\rangle$,  the medians being computed among  all mass bins
and galactocentric distances. The  values of both $\langle g-X\rangle$
and  $OFF_X$,  for SSP  models  from  different  synthesis codes,  are
reported  in Tab.~\ref{tab:colours_median_offsets}.  Dotted  curves in
Fig.~\ref{fig:col_profs_offsets} are  obtained for CB07SC  SSP models,
by  minimizing Eq.~\ref{eq:chi}  with  the values  of  $OFF_X$ set  to
zero.   Notice  the  significant   difference  between   observed  and
(best-fitting)  model  colours,  ranging  from  a  few  hundredths  of
magnitude ($g-r$ and  $g-K$) up to $\sim 0.1$~mag  in $g-Y$. To derive
the offsets,  $OFF_X$, we set  $OFF_X=0$ in Eq.~\ref{eq:chi},  and fit
the observed colours with a given  set of models (e.g. SSPs), for each
galaxy mass  bin and galactocentric  distance.  The values  of $OFF_X$
are defined as the median differences of all observed and best-fitting
colours. After  including the  $OFF_X$ terms in  Eq.~\ref{eq:chi}, and
repeating  the  minimization  procedure,   we  find  that  the  median
differences of  observed and best-fitting colours  become smaller than
$\sim   0.007$~mag,    proving   that   the    entire   procedure   is
self-consistent.  We note that by  using $\tau$ and burst (rather than
simple) SP models, the offsets change less than $\sim 0.01$~mag, while
the offsets  differ by $\simlt  0.03$~mag for SSPs among  different SP
codes (see  Tab.~\ref{tab:colours_median_offsets}).  The dashed curves
in  Fig.~\ref{fig:col_profs_offsets} show  the  offset-corrected model
colours,  i.e. $(g-X)_{MOD}-OFF_X$,  hereafter simply  referred  to as
best-fitting model  colours.  Notice how the matching  of observed and
model colours improves significantly when applying the offsets.
Fig.~\ref{fig:col_profs_fits} plots, as an example, the observed
colours at all fiducial galactocentric distances (see
Sec.~\ref{sec:profiles}), for low-, intermediate-, and high-mass
galaxies, and corresponding best-fitting colours for CB$^\ast \!$MC SSPs.
Overall,  we  find  that  all  the adopted  models  (i.e.   population
synthesis codes,  and star formation histories) provide  a fairly good
match to the observations in  all available passbands (but for J band,
see above).   { We  characterize the quality  of colour  fitting by
  computing the  median rms value of differences  between best-fit and
  observed  colours,  the median  being  computed  among all  fiducial
  radii.   These  rms  values  are  summarized, for  all  colour  fits
  performed in  the present work,  in Tab.~\ref{tab:rms_fits}.  Notice
  that the typical rms amounts  to $\sim 0.02$~mag.  For what concerns
  best-fitting age and metallicity, averaging over all radial bins and
  all  models, the (absolute)  values of  age (metallicity)  differ by
  $\sim 18  \%$ ($14\%$) when (i)  one applies the  $OFF_X$'s and (ii)
  the $OFF_X$'s are  set to zero.  On the  other hand, relative values
  of    age    and    metallicity    (i.e.    the    gradients,    see
  Sec.~\ref{sec:SP_Age_Z}) are virtually  unaffected by the $OFF_X$'s.
  In  order   to  estimate  uncertainties  on   best-fitting  age  and
  metallicity,  we  perform $200$  bootstrap  iterations, where,  each
  time, observed colours are shifted according to their errors and the
  entire minimization procedure is repeated.  This bootstrap procedure
  provides a joint probability distribution function (PDF) for age and
  metallicity.   Confidence  intervals  on  age  and  metallicity  are
  obtained at  the 1~$\sigma$ equivalent,  by computing the  16-th and
  84-th percentiles of the marginalized PDF.
}

\begin{table*}
\centering
\begin{minipage}{180mm}
\small
\centering
\caption{{ Median rms of  differences between best-fit and observed
    colours, for  all models  and samples analyzed  in this  work. The
    median values are computed among all fiducial radii, and inform on
    the  quality of  the colour  fitting.  For  CB$^\ast$  models, the
    offsets corresponding to a Salpeter (S), rather than Chabrier (C),
    IMF are given in brackets.}}
\begin{tabular}{c|c|c|c|c}
\hline
model & \multicolumn{3}{c}{median rms}\\
\hline
             &     low- & intermediate- & high-mass \\
\hline
BC03 SSPs                          &  $0.017$ & $0.015$ & $0.008$ \\ 
CB07SC SSPs                        &  $0.016$ & $0.020$ & $0.013$ \\ 
CB$^\ast \!$IC (CB$^\ast \!$IS) SSPs &  $0.022$($0.020$) & $0.017$($0.012$) & $0.019$($0.019$) \\ 
CB$^\ast \!$MC (CB$^\ast \!$MS) SSPs &  $0.032$($0.016$) & $0.010$($0.010$) & $0.015$($0.015$) \\ 
CB$^\ast \!$SC (CB$^\ast \!$SS) SSPs &  $0.015$($0.015$) & $0.014$($0.015$) & $0.014$($0.012$) \\ 
CB$^\ast \!$BC (CB$^\ast \!$BS) SSPs &  $0.019$($0.016$) & $0.013$($0.011$) & $0.024$($0.022$) \\ 
CB$^\ast \!$SC $\tau$ models        &  $0.016$ & $0.015$ & $0.014$ \\ 
CB$^\ast \!$SC burst models         &  $0.016$ & $0.015$ & $0.013$ \\ 
CB07SC ($E(B-V)=0.04$)             &  $0.026$ & $0.024$ & $0.021$ \\ 
CB07SC ($E(B-V)=0.08$)             &  $0.023$ & $0.019$ & $0.020$ \\ 
CB07SC (for field ETGs)            &  $0.020$ & $0.019$ & $0.026$ \\ 
CB$^\ast \!$MC (for field ETGs)     &  $0.015$ & $0.017$ & $0.024$ \\ 
CB07SC (for group ETGs)            &  $0.019$ & $0.021$ & $0.020$ \\ 
CB$^\ast \!$MC (for group ETGs)     &  $0.012$ & $0.020$ & $0.017$ \\ 
\hline
\end{tabular}
\label{tab:rms_fits}
\end{minipage}
\end{table*}



\begin{table*}
\centering
\begin{minipage}{180mm}
\small
\centering
\caption{Median colours of ETGs and photometric offsets, $OFF_X$,
  applied to different SSP models. For CB$^\ast$ models, the offsets
  corresponding to a Salpeter (S), rather than Chabrier (C), IMF are
  given in brackets.}
\begin{tabular}{c|c|r|r|r|r|r|r}
\hline
 $colour$ & $median$ & \multicolumn{6}{c}{$OFF_X$}\\
\hline
 & & BC03 & CB07SC& CB$^{\ast \!}$IC(S) &CB$^{\ast \!}$MC(S) &CB$^{\ast \!}$SC(S) &CB$^{\ast \!}$BC(S) \\
\hline
\hline 
$g-r$   & $0.803$ &$ -0.033$ &$ -0.019$ &$ -0.028$$( -0.027)$ &$ -0.022$$( -0.022)$ &$ -0.025$$( -0.022)$ &$ -0.016$$( -0.018)$ \\
$g-i$   & $1.197$ &$  0.032$ &$  0.031$ &$  0.028$$(  0.029)$ &$  0.035$$(  0.034)$ &$  0.024$$(  0.023)$ &$  0.039$$(  0.039)$ \\
$g-z$   & $1.458$ &$  0.044$ &$  0.016$ &$  0.054$$(  0.052)$ &$  0.042$$(  0.034)$ &$  0.049$$(  0.044)$ &$  0.023$$(  0.018)$ \\
$g-Y$   & $2.152$ &$ -0.119$ &$ -0.131$ &$ -0.109$$( -0.114)$ &$ -0.129$$( -0.132)$ &$ -0.117$$( -0.121)$ &$ -0.131$$( -0.131)$ \\
$g-H$   & $3.286$ &$  0.035$ &$  0.060$ &$  0.016$$(  0.018)$ &$  0.033$$(  0.040)$ &$  0.028$$(  0.032)$ &$  0.044$$(  0.045)$ \\
$g-K$   & $3.627$ &$  0.037$ &$  0.041$ &$  0.014$$( -0.011)$ &$  0.017$$(  0.027)$ &$  0.017$$(  0.024)$ &$  0.025$$(  0.034)$ \\
\hline
\end{tabular}
\label{tab:colours_median_offsets}
\end{minipage}
\end{table*}

\section{Age and metallicity gradients out to $8\,{\rm R}_e$}
\label{sec:SP_Age_Z}

\subsection{SSP models}
\label{subsec:SSP_gradients}
\begin{figure*}
\begin{center}
\includegraphics[width=170mm]{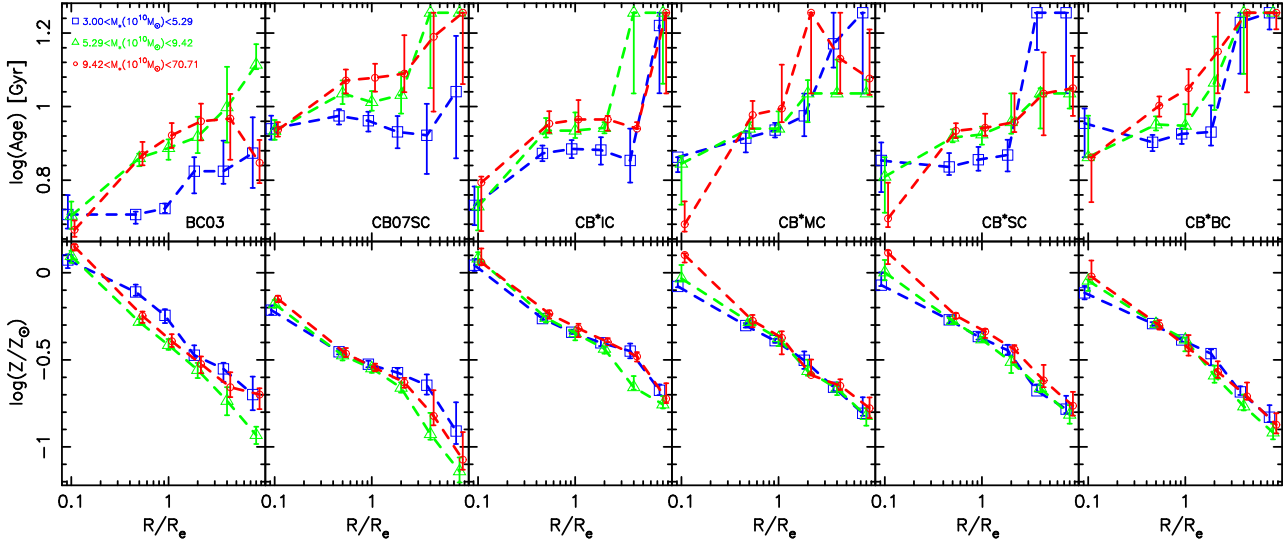}
\end{center}
\caption{Age and metallicity profiles obtained by fitting galaxy
  colours with SSP models, as a function of galactocentric distance,
  for low-(blue), intermediate-(green), and high-(red) mass ETGs.
  Different symbols are used for different galaxy mass bins, as
  illustrated in the top-left panel. From left to right, top and
  bottom panels refer to different models (BC03, CB07SC, CB$^\ast \!$IC,
  CB$^\ast \!$MC, CB$^\ast \!$SC, CB$^\ast \!$BC), as illustrated by the black labels in the
  top panels. All models correspond to a Chabrier IMF. Error bars
  denote the 16$th$ and 84$th$ percentile uncertainties (corresponding
  to $1\sigma$ errors for a normal deviate) on best-fitting age and
  metallicity.  }
\label{fig:agemet_profs}
\end{figure*}

\begin{figure*}
\begin{center}
\includegraphics[width=120mm]{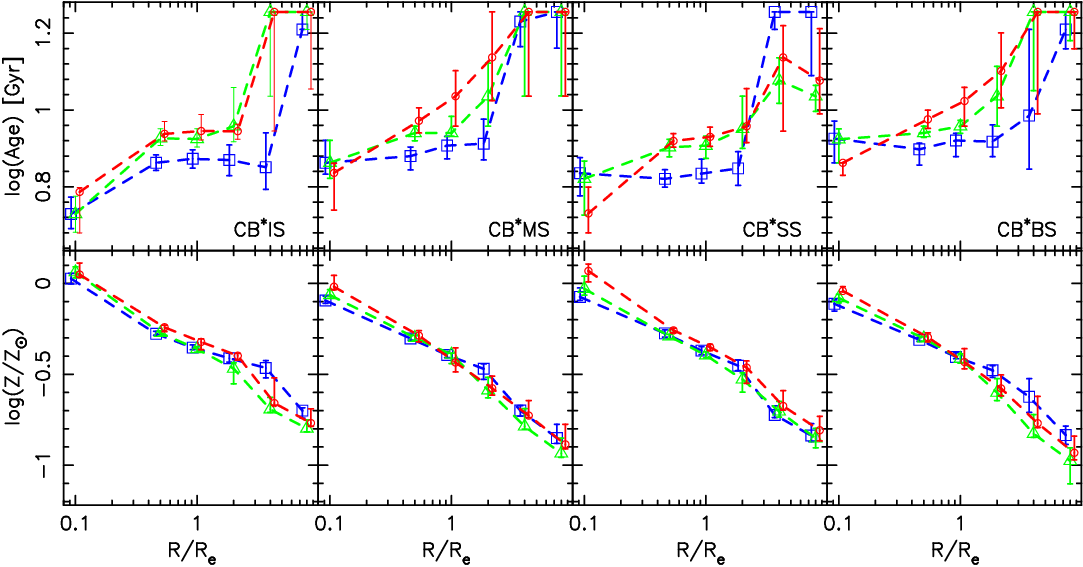}
\end{center}
\caption{Same as Fig.~\ref{fig:agemet_profs} but plotting only the
  results for the CB$^\ast$ models using a Salpeter IMF.}
\label{fig:agemet_profs_SALP}
\end{figure*}

Fig.~\ref{fig:agemet_profs} plots the best-fitting age and metallicity
as  a  function  of  galactocentric  distance,  for  BC03,  CB07,  and
CB$^\ast$ (Chabrier IMF) SSP models. In the radial range from $0.1$ to
$1\,\rm R_e$,  the $Age$  parameter (i.e. the  ``SSP-equivalent'' age)
increases,  while  metallicity  decreases,  implying  that  ETGs  have
negative metallicity gradients  and positive age gradients, consistent
with                         previous                         findings
\citep[e.g.][]{McClure69,PVJ90,gorgas90,GON93,davies93,ferr05,ferr09,PSB:06,LdC:09,Clemens:09,SUH:10,Tor:10}.
In the present  study, we supersede previous works,  by exploring, for
the  first  time,  the  behaviour  of age  and  metallicity  at  large
galactocentric distances. Overall, we  find that at $\rm R>R_e$, $Age$
keeps  increasing, while  metallicity decreases,  out to  $\sim 8\,\rm
R_e$. Notice that the error bars  on $Age$ are on average large (up to
$40 \%$ for some models and  mass bins) at $\rm R>R_e$, reflecting the
uncertainties  on galaxy  colours, as  well as  the fact  that ``old''
(i.e.   $\simgt  $4~Gyr)  ages  are intrinsically  more  difficult  to
constrain.   { This  is due  to the  fact that  for an  old stellar
  population,  a variation  in  age  does not  change  much the  model
  colors.  Hence, the  large uncertainties on $Age$ at  $\rm R>R_e$ do
  not  affect   significantly  those  in   metallicity,  which  remain
  reasonably  small  ($<10$--$20\%$)   even  at  large  galactocentric
  distances.  } Furthermore, the radial trend with metallicity appears
quite  robust   across  the  different   population  synthesis  models
considered.    Fig.~\ref{fig:agemet_profs_SALP}  shows  the   age  and
metallicity  profiles  for  CB$^\ast$  models when  using  a  Salpeter
IMF.  Overall, the  profiles are  very similar  to those  presented in
Fig.~\ref{fig:agemet_profs}, implying  that the  choice of IMF  in the
stellar population models does not change at all our conclusions.



In  order  to  quantify the  change  in  the  slopes  of the  age  and
metallicity  profiles,   i.e.   the  age   and  metallicity  gradients
(hereafter \nablat\ and \nablaz), as  a function of radius, we perform
an  orthogonal  linear  regression   of  the  profiles  in  the  inner
($\rm\log(R/R_e)\leq   0$)  and  outer   ($\rm\log(R/R_e)>0$)  regions
separately. The slopes give  the inner and outer metallicity gradients
(\nablazi\ and \nablazo), and the inner age gradient (\nablati). Since
age profiles become noisier at $\rm  R>R_e$, we do not perform any fit
of  the $Age$  profile  in  the outskirts.   Instead,  we compute  the
average,  logarithmic difference  in  $Age$, \dlt,  between the  outer
region~\footnote{ In  practice, we compute the mean  value of $Age$
  among the  three fiducial radii of  2, 4, and 8~$\rm  R_e$, and then
  take  the logarithmic  difference between  this mean  value  and the
  $Age$ at  $\rm R=0.1  \, R_e$.}  ($\rm R >  1\,R_e$) and  the galaxy
centre  ($\rm  R=0.1  \,  R_e$).   { On  the  contrary,  errors  on
  metallicity remain small even at large galactocentric distances (see
  above),  allowing  us   to  estimate  meaningful  outer  metallicity
  gradients.  Errors  on \nablat, \nablaz, and \dlt\  are estimated as
  the    errors    on    age    and   metallicity    (see    end    of
  Sec.~\ref{subsec:color_fitting}),    performing    $200$   bootstrap
  iterations,   where   colours  are   shifted   according  to   their
  uncertainties, and  age and metallicity  gradients are re-estimated,
  by  repeating the  $\chi^2$ minimization  procedure at  each fiducal
  radius}.  For all different SSP  models, the values of \nablazi\ and
\nablazo\  (\nablati\  and  \dlt)  are reported  in  Tab.~\ref{tab:nz}
(Tab.~\ref{tab:nt}).

Fig.~\ref{fig:agemet_profs}   unsurprisingly   shows   that   absolute
estimates  of   $Age$  and  metallicity   differ  significantly  among
different models,  while {\it relative}  trends, i.e. with  respect to
mass  and galactocentric  distance, do  not  (but see  below). In  the
galaxy  centre   ($\rm  R\sim   0.1\,R_e$),  BC03  models   predict  a
super-solar  metallicity ($\sim 1.4  Z_\odot$) and  an $Age$  of $\sim
5$~Gyr, while  for CB07, the  central metallicity is  sub-solar ($\sim
0.65  Z_\odot$), and  $Age$  older  (with respect  to  BC03) by  $\sim
4$~Gyr.  The new CB$^\ast$  models also exhibit significant reciprocal
differences  in  $Age$  (up  to  a few  Gyrs),  and  relatively  small
differences in  $Z$ ($\sim 0.1$~dex), when comparing  results based on
different stellar libraries  (CB$^\ast \!$IC; CB$^\ast \!$MC; CB$^\ast
\!$SC; CB$^\ast \!$BC). Notice that, for high-mass ETGs, all CB$^\ast$
models  predict  either  solar  (CB$^\ast \!$MC;  CB$^\ast  \!$BC)  or
super-solar (CB$^\ast \!$IC;  CB$^\ast \!$SC) metallicities. Hence, in
contrast  to  the CB07  models,  the  photometric  constraints of  the
CB$^\ast$   models  are   more  consistent   with  the   results  from
spectroscopic    studies   of    the   central    regions    of   ETGs
(e.g.~\citealt{GALL:06,  Pasquali:10}).  Since  the absolute  $Age$ is
strongly model dependent (even  when changing only the adopted stellar
library in CB$^\ast$),  we have not set the age of  the Universe as an
upper  limit  for  the  models (see  Sec.~\ref{subsec:SPmodels}),  and
several   profiles  in  Fig.~\ref{fig:agemet_profs}   keep  increasing
outwards  up to  the maximum  allowed value  of $18$~Gyr.  Imposing an
$Age$ upper limit of $13.75$~Gyr \citep[i.e. the current best estimate
  for the age of the Universe,][]{wmap7} would produce an upper cutoff
in  the   $Age$  profiles,  while  not   affecting  significantly  the
metallicity  profiles. Considering the  large $Age$  differences among
different models, enforcing an upper  limit on $Age$ is not meaningful
for the models, so we decided not to apply this constraint.

Overall, the key features of the age and metallicity
trends are summarised as follows:
\begin{description}
\item[-] Metallicity is a monotonic, decreasing function of
  galactocentric distance in ETGs, out to (at least) $8\,\rm R_e$. This
  is the most robust result, holding true for all mass bins, and for
  all different models probed here.
\item[-] The values of \dlt\ in Tab.~\ref{tab:nt} are always
  positive. In most cases (and in particular for all CB$^\ast$ models with
  Chabrier IMF), this result is found above the $2\sigma$ level. For
  high-mass galaxies, the \dlt\ is positive at more than the $3\sigma$
  level for all models. We conclude that ETGs, in particular the most
  massive ones, host significantly older stellar populations in their
  halos ($\rm R\simgt few\times R_e$) than in their cores.
\item[-]  In the  inner  region, the  metallicity gradient,  \nablazi,
  steepens  at high  mass, while  \nablati\ increases  with $M_\star$.
  These trends  hold for all  models, except for  CB$^\ast \!$/IndoUS,
  where no mass dependence is  detected. A steepening of \nablazi\ and
  \nablati\ with  galaxy mass has  also been found in  previous papers
  based on  the SPIDER survey (e.g.~Paper  IV; \citealt{LF:11}), where
  colour  gradients   were  fitted  (rather  than   colours)  with  SP
  models. Notice that at high  mass, the \nablati\ is always positive,
  regardless  of  the  SP  model,  consistent with  our  findings  for
  \dlt\ in  the external regions.  At intermediate-  and low-mass, the
  \nablat\  is either  consistent  with zero,  within  the errors,  or
  significantly positive.   { In particular,  at low mass,  the age
    trend is flat for most models (i.e.  BC03, CB07SC, CB$^\ast \!$SC,
    and CB$^\ast \!$BC). }
\item[-] For low-mass ETGs, the metallicity gradient is steeper at
  large radii than in the centre, i.e. \nablazo$<$\nablazi.  While
  this result holds for all models, it is statistically significant
  (at more than a 2--3$\sigma$ level) only for some models
  (e.g.CB$^\ast \!$SC). At high-mass, the same steepening is observed only for
  some models when using a Chabrier IMF (e.g. CB$^\ast \!$IC), and for all
  models with a Salpeter IMF (in which case, however, the steepening
  is statistically significant only for CB$^\ast \!$IS models). If a
  bottom-heavy, Salpeter-like, IMF is more appropriate for high-,
  relative to low-mass ETGs (e.g.~\citealt{vDC:11, Cappellari:12}), we
  might be tempted to conclude that the metallicity gradient of ETGs
  steepens at large galacto-centric distances, independent of galaxy
  stellar mass.
\end{description}

\subsection{Exponential and burst models}

\begin{figure}
\begin{center}
\includegraphics[width=70mm]{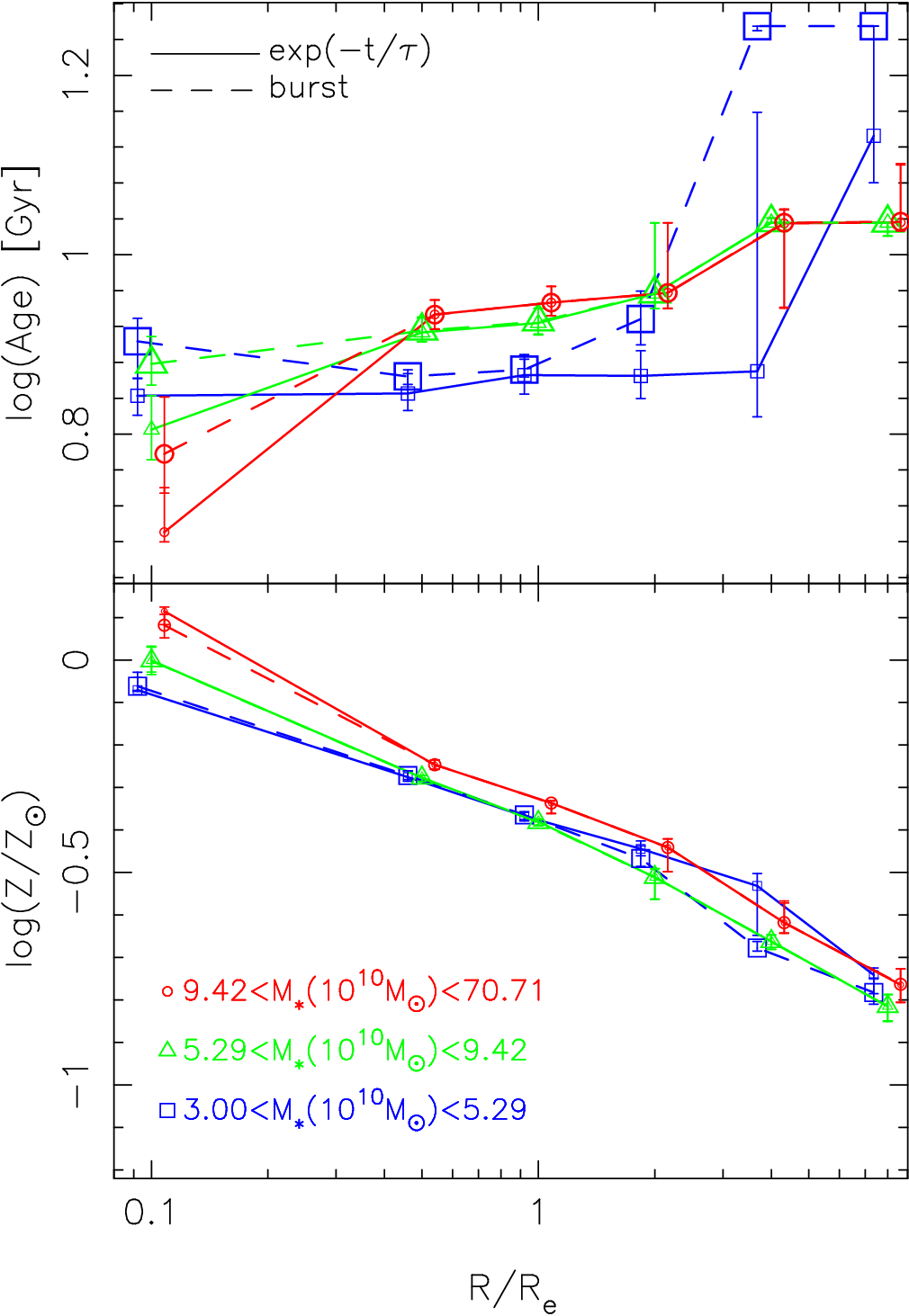}
\end{center}
\caption{The best-fitting age (top) and metallicity (bottom) are shown
  as a function of galactocentric distance, as in
  Fig.~\ref{fig:agemet_profs}, for low-(blue), intermediate-(green),
  and high-(red) mass ETGs. Smaller (larger) symbols and solid
  (dashed) curves refer to CB$^\ast \!$SC exponentially declining (burst)
  models.  Notice that the trends are very similar to those obtained
  with SSP models (Fig.~\ref{fig:agemet_profs}).  }
\label{fig:agemet_exp_burst}
\end{figure}

Fig.~\ref{fig:agemet_exp_burst} shows radial trends of (mass-weighted)
age  and  metallicity  as  obtained  by fitting  galaxy  colours  with
composite stellar  populations ($\tau$ and burst  models). For brevity
reasons, we only show one illustrative case, corresponding to CB$^\ast
\!$SC   models  (Sec.~\ref{subsec:color_fitting}).   Both   $Age$  and
metallicity  exhibit  similar  trends  to  those  obtained  with  SSPs
(Fig.~\ref{fig:agemet_profs}).  In  particular, we see  that (i) $Age$
increases outwards, being  oldest in the galaxy halo  region; (ii) the
age gradient is  stronger (more positive) at higher  galaxy mass{ ,
  with low-mass  ETGs having a flat  age trend out  to $\sim $2--4$\rm
  R_e$};  and (iii)  metallicity  decreases outwards,  with a  steeper
(more  negative) gradient (especially  for low-  and intermediate-mass
galaxies) at large galactocentric distances.  One may also notice that
individual values of $Age$ are somewhat more sensitive (in contrast to
metallicity) to the adopted star formation history (i.e.  SSP, $\tau$,
and burst models).  For instance,  at $\rm R>3\,R_e$, and for low-mass
ETGs, both burst and SSP models (see Fig.~\ref{fig:agemet_profs}) give
the oldest  available $Age$  (18~$Gyr$) in the  input grid  of values,
while for $\tau$ models the $Age$ is $\sim 13$~Gyr (see blue curves in
the upper  panel of the Figure).  On the other hand,  all models agree
very well (within error bars)  on the inferred value of metallicity at
each radial position.

\subsection{Internal reddening}

\begin{figure}
\begin{center}
\includegraphics[width=70mm]{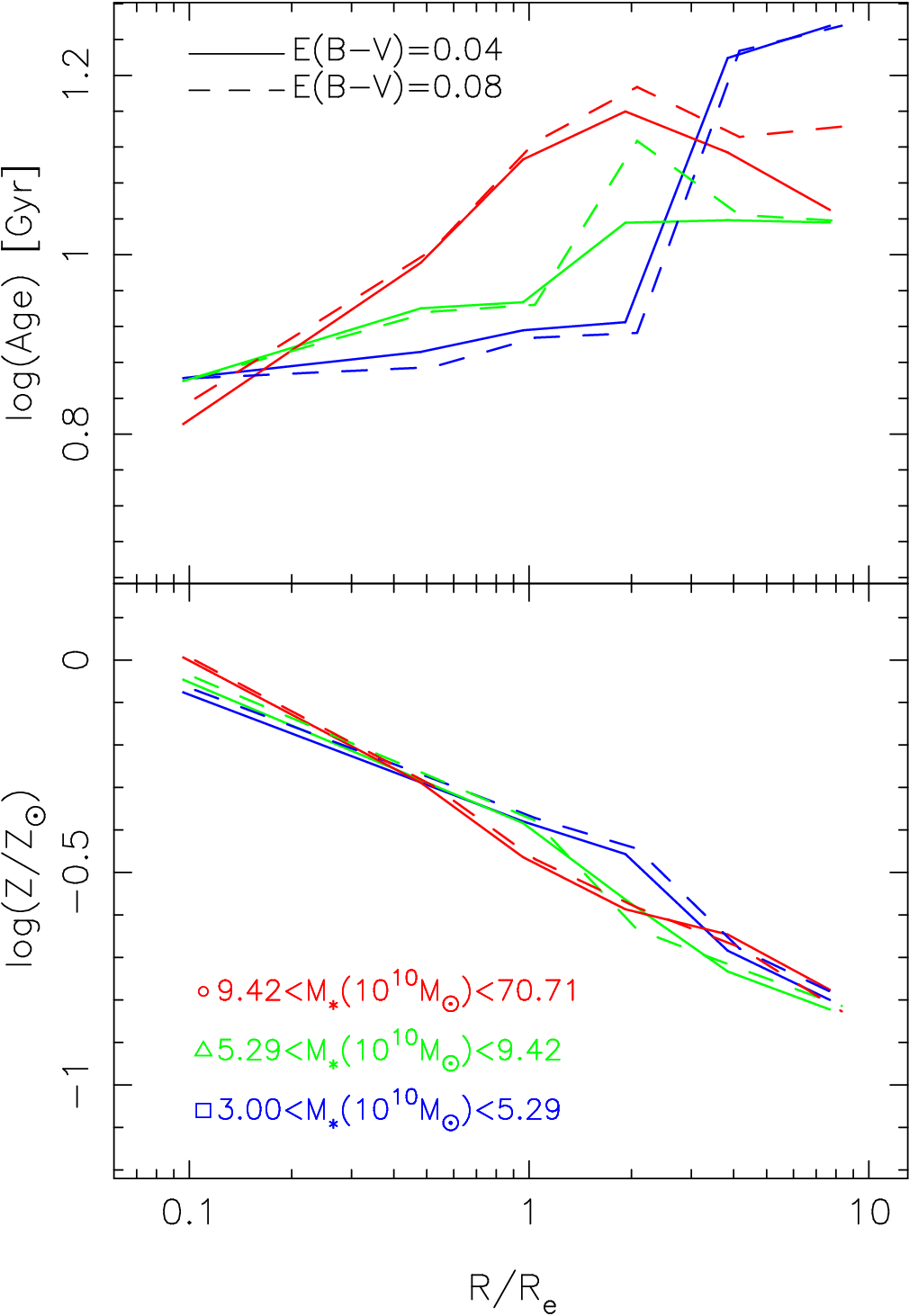}
\end{center}
\caption{The same as right panel of Fig.~\ref{fig:agemet_profs}, but
  comparing results of CB07SC with ~\citet{Cardelli} extinction law
  applied, with a colour excess of $E(B-V)=0.04$ (solid curves) and
  $E(B-V)=0.08$ (dashed curves). Notice that internal reddening does
  not affect significantly the trends of age and metallicity with
  galactocentric distance.}
\label{fig:EBV_age_met}
\end{figure}

So far, we have neglected the role of internal reddening, which might
mimic the presence of metallicity gradients in ETGs~\citep{GJON95,
  SW:96, Wise:96}. Indeed, if dust gradients are present, a
correlation of colour gradients and the amount of internal extinction
should be detected. Such correlation does not exist but for a small
fraction ($< 10 \%$) of ETGs in our sample (see Paper IV). Also,
previous studies have found neither a significant correlation between
colour gradients and IRAS 100~$\mu$ flux~\citep{MIC:05}, nor a
contribution of diffuse dust to the colour
gradients~\citep{SWF:09}. Although the evidence supports our
assumption of neglecting extinction gradients in ETGs, the presence of
uniform internal reddening inside the galaxies may still affect our
conclusions, as it could bias the best-fitting values of age and
metallicity at each radial position.  To address this issue, we have
applied dust extinction corrections to one of the available sets of
CB$^\ast$ SSPs (CB$^\ast \!$MC), adopting the~\citet{Cardelli} extinction law, and
repeated the SSP fitting procedure for two cases, i.e. a colour excess
of $E(B-V)=0.04$ and $E(B-V)=0.08$. These values correspond to the
median and the 75\% level of the probability distribution of the
colour excess parameter, inferred by running the spectral fitting code
STARLIGHT~\citep{CID05}, with the CB07SC models, on the SDSS spectra
of our sample of ETGs (see~Paper V for details).
Fig.~\ref{fig:EBV_age_met} shows the age and metallicity profiles
obtained when adopting the above reddening values. The trends of age
and metallicity are fully consistent with those obtained for the
dustless models (see CB$^\ast \!$MC profiles in Fig.~\ref{fig:agemet_profs}).

\begin{table*}
\centering
\small
\caption{Metallicity gradients (\nablaz) of ETGs estimated using
  different SSP models.  Subscript $i$ and $o$ refer to the inner
  ($\rm R \le R_e$) and outer ($\rm R_e<R \le 8\,R_e$) radial ranges
  where the gradients are estimated.}
\begin{tabular}{c|c|c|c|c|c|c|c}
\hline
model & \multicolumn{7}{c}{METALLICITY GRADIENTS} \\
\hline
       & \multicolumn{3}{c}{$\nabla_{Z,i} $} & & \multicolumn{3}{c}{$\nabla_{Z,o} $} \\ 
\cline{2-4} \cline{6-8} 
       & low- & intermediate- & high-mass & & low- & intermediate- & high-mass \\ 
BC03   & $ -0.31\pm0.06$&$ -0.52\pm0.05$&$ -0.55\pm0.04$& &$ -0.38\pm0.17$&$ -0.62\pm0.12$&$ -0.29\pm0.12$\\
CB07SC & $ -0.32\pm0.04$&$ -0.38\pm0.03$&$ -0.41\pm0.03$& &$ -0.57\pm0.26$&$ -0.80\pm0.13$&$ -0.74\pm0.20$\\
CB$^\ast \!$IC & $ -0.40\pm0.04$&$ -0.44\pm0.05$&$ -0.39\pm0.05$& &$ -0.46\pm0.06$&$ -0.54\pm0.07$&$ -0.55\pm0.12$\\
CB$^\ast \!$MC & $ -0.30\pm0.04$&$ -0.35\pm0.05$&$ -0.49\pm0.07$& &$ -0.58\pm0.14$&$ -0.43\pm0.10$&$ -0.32\pm0.13$\\
CB$^\ast \!$SC & $ -0.29\pm0.05$&$ -0.39\pm0.07$&$ -0.46\pm0.05$& &$ -0.57\pm0.09$&$ -0.50\pm0.12$&$ -0.54\pm0.15$\\
CB$^\ast \!$BC & $ -0.27\pm0.03$&$ -0.34\pm0.06$&$ -0.41\pm0.07$& &$ -0.61\pm0.12$&$ -0.54\pm0.11$&$ -0.51\pm0.13$\\
CB$^\ast \!$IS & $ -0.39\pm0.03$&$ -0.43\pm0.04$&$ -0.38\pm0.06$& &$ -0.49\pm0.07$&$ -0.55\pm0.13$&$ -0.62\pm0.08$\\
CB$^\ast \!$MS & $ -0.30\pm0.03$&$ -0.33\pm0.03$&$ -0.41\pm0.08$& &$ -0.64\pm0.13$&$ -0.58\pm0.13$&$ -0.51\pm0.13$\\
CB$^\ast \!$SS & $ -0.29\pm0.05$&$ -0.37\pm0.05$&$ -0.43\pm0.06$& &$ -0.65\pm0.10$&$ -0.54\pm0.13$&$ -0.58\pm0.16$\\
CB$^\ast \!$BS & $ -0.29\pm0.03$&$ -0.33\pm0.03$&$ -0.39\pm0.06$& &$ -0.59\pm0.11$&$ -0.63\pm0.20$&$ -0.59\pm0.16$\\
\hline
\end{tabular}
\label{tab:nz}
\end{table*}

\begin{table*}
\centering
\small
\caption{Same as Tab.~\ref{tab:nz} but for age, rather than
  metallicity, gradients (\nablat).  Notice that instead of computing
  age gradients in the galaxy outskirts, we measure the mean
  logarithmic age difference, \dlt, between the outer radial range and
  the galaxy centre (see text for details).}
\begin{tabular}{c|c|c|c|c|c|c|c}
\hline
model & \multicolumn{7}{c}{AGE GRADIENTS} \\
\hline
       & \multicolumn{3}{c}{$\nabla_{t,i} $} & & \multicolumn{3}{c}{\dlt} \\ 
\cline{2-4} \cline{6-8} 
       & low- & intermediate- & high-mass & & low- & intermediate- & high-mass \\ 
BC03   & $  0.01\pm0.03$&$  0.19\pm0.03$&$  0.26\pm0.02$&&$  0.13\pm0.06$&$  0.31\pm0.05$&$  0.26\pm0.04$\\
CB07SC & $  0.03\pm0.03$&$  0.09\pm0.02$&$  0.15\pm0.02$&&$  0.03\pm0.08$&$  0.25\pm0.05$&$  0.24\pm0.07$\\
CB$^\ast \!$IC & $  0.16\pm0.03$&$  0.22\pm0.04$&$  0.18\pm0.06$&&$  0.25\pm0.08$&$  0.42\pm0.08$&$  0.26\pm0.10$\\
CB$^\ast \!$MC & $  0.04\pm0.04$&$  0.10\pm0.05$&$  0.34\pm0.05$&&$  0.28\pm0.07$&$  0.19\pm0.09$&$  0.47\pm0.08$\\
CB$^\ast \!$SC & $ -0.00\pm0.05$&$  0.12\pm0.06$&$  0.26\pm0.05$&&$  0.27\pm0.09$&$  0.20\pm0.09$&$  0.32\pm0.09$\\
CB$^\ast \!$BC & $ -0.04\pm0.05$&$  0.09\pm0.05$&$  0.19\pm0.06$&&$  0.18\pm0.08$&$  0.33\pm0.08$&$  0.36\pm0.10$\\
CB$^\ast \!$IS & $  0.15\pm0.05$&$  0.21\pm0.05$&$  0.17\pm0.09$&&$  0.25\pm0.05$&$  0.43\pm0.07$&$  0.37\pm0.10$\\
CB$^\ast \!$MS & $  0.04\pm0.06$&$  0.08\pm0.06$&$  0.20\pm0.09$&&$  0.27\pm0.06$&$  0.32\pm0.09$&$  0.38\pm0.10$\\
CB$^\ast \!$SS & $ -0.00\pm0.06$&$  0.09\pm0.08$&$  0.21\pm0.07$&&$  0.28\pm0.06$&$  0.20\pm0.08$&$  0.33\pm0.09$\\
CB$^\ast \!$BS & $ -0.01\pm0.06$&$  0.03\pm0.05$&$  0.16\pm0.05$&&$  0.11\pm0.09$&$  0.26\pm0.07$&$  0.34\pm0.07$\\
\hline
\end{tabular}
\label{tab:nt}
\end{table*}

\begin{table*}
\centering
\small
\caption{Metallicity (\nablaz) and age (\nablat) gradients of ETGs as
  a function of environment, for CB07SC SSP models. The subscripts $i$ and
  $o$ refer to the inner ($\rm R\le R_e$) and outer ($\rm R_e<R<8\,R_e$)
  radial ranges where gradients are estimated.}
\begin{tabular}{c|c|c|c|c}
\hline
mass range & $\nabla_{Z,i} $ & $\nabla_{Z,o} $ & $\nabla_{t,i} $ & \dlt \\ 
\hline
\multicolumn{5}{c}{FIELD ETGs} \\
$3.00 < M_{\star}[10^{10} M_{\odot}]<5.29$  & $-0.32\pm 0.04$ & $-0.87\pm  0.25$&$ -0.01\pm  0.06$&$ -0.02\pm  0.11$\\
$5.29 < M_{\star}[10^{10} M_{\odot}]< 9.42$ & $-0.35\pm 0.04$ & $-0.13\pm  0.17$&$  0.05\pm  0.05$&$ 0.01\pm  0.11$\\
$9.42 < M_{\star}[10^{10} M_{\odot}]<70.71$ & $-0.39\pm 0.03$ & $-0.15\pm  0.22$&$  0.06\pm  0.05$&$ 0.06\pm  0.10$\\
\hline
\multicolumn{5}{c}{GROUP ETGs} \\
$3.00 < M_{\star} [10^{10} M_{\odot}] <5.29$&$ -0.36\pm  0.04$&$ -0.47\pm  0.18$&$  0.08\pm  0.05$&$  0.10\pm  0.08$\\
$5.29 < M_{\star} [10^{10} M_{\odot}] < 9.42$&$ -0.36\pm  0.03$&$ -0.92\pm  0.12$&$  0.10\pm  0.04$&$  0.25\pm  0.04$\\
$9.42 < M_{\star} [10^{10} M_{\odot}] <70.71$&$ -0.41\pm  0.03$&$ -0.94\pm  0.25$&$  0.20\pm  0.04$&$  0.30\pm  0.10$\\
\hline
\end{tabular}
\label{tab:nz_nt_env}
\end{table*}


\begin{table*}
\centering
\small
\caption{The same as Tab.~\ref{tab:nz_nt_env} but for CB$^\ast \!$MC rather than CB07SC SSP models.}
\begin{tabular}{c|c|c|c|c}
\hline
mass range & $\nabla_{Z,i} $ & $\nabla_{Z,o} $ & $\nabla_{t,i} $ & \dlt \\ 
\hline
\multicolumn{5}{c}{FIELD ETGs} \\
$3.00 < M_{\star}[10^{10} M_{\odot}]<5.29$  & $ -0.37\pm  0.06$&$ -0.87\pm  0.30$&$  0.09\pm  0.08$&$  0.38\pm  0.09$\\
$5.29 < M_{\star}[10^{10} M_{\odot}]< 9.42$ & $ -0.34\pm  0.08$&$ -0.36\pm  0.16$&$  0.12\pm  0.13$&$  0.21\pm  0.13$ \\
$9.42 < M_{\star}[10^{10} M_{\odot}]<70.71$ & $ -0.37\pm  0.07$&$ -0.15\pm  0.16$&$  0.11\pm  0.08$&$  0.16\pm  0.06$\\
\hline
\multicolumn{5}{c}{GROUP ETGs} \\
$3.00 < M_{\star} [10^{10} M_{\odot}] <5.29$& $ -0.39\pm  0.06$&$ -0.57\pm  0.11$&$  0.22\pm  0.10$&$  0.43\pm  0.10$\\
$5.29 < M_{\star} [10^{10} M_{\odot}] < 9.42$&$ -0.32\pm  0.03$&$ -0.56\pm  0.10$&$  0.11\pm  0.03$&$  0.19\pm  0.03$\\
$9.42 < M_{\star} [10^{10} M_{\odot}] <70.71$&$ -0.58\pm  0.04$&$ -0.52\pm  0.13$&$  0.52\pm  0.06$&$  0.54\pm  0.06$\\
\hline
\end{tabular}
\label{tab:nz_nt_env_CB13}
\end{table*}



\subsection{Comparison with recent findings}

Our findings (e.g.~Fig.~\ref{fig:agemet_profs}) can be compared with
those of~\citet[hereafter CGA10]{CGA:10}, who derived radial profiles
of stellar population properties (i.e. SSP-equivalent age, [Z/H], and
[$\alpha$/Fe]), based on spectroscopic data, for one of the two
brightest galaxies in the Coma cluster ($NGC \, 4889$), out to a
galactocentric distance of $\sim 4\,\rm R_e$. CGA10 found that the stellar
populations in the outer halo of $NGC\,4889$ tend to be older than
those in the inner regions, in agreement with our results. At $\rm R\simgt
1.2\,R_e$, the metallicity profile of $NGC\,4889$ flattens, while
[$\alpha$/Fe] shows a steep negative gradient. Although the
optical+NIR photometry allows us to reasonably disentangle the average
contribution of age and metallicity to galaxy colours (see Paper IV),
we are not able to constrain both [Z/H] and [$\alpha$/Fe] separately,
and our metallicity profiles can indeed be affected by radial changes
of [$\alpha$/Fe].


For instance, the steepening of the metallicity profile at $\rm R>R_e$
in Figs.~\ref{fig:agemet_profs} and~\ref{fig:agemet_exp_burst} may be
due to a steepening of the [$\alpha$/Fe] profile at large
galactocentric distances, as observed by CGA10, or, alternatively, it
can be a true change in [Z/H], implying that the result of CGA10 does
not apply to the entire population of ETGs in our sample. The fact
that we obtain similar trends of $Age$ and [Z/H] for both $\tau$ and
SSP models favours a true variation of the metallicity profile in the
galaxy outskirts rather than a variation of the star formation
time-scale (i.e. abundance ratios) at large radii. In this regard, our
results are more consistent with those recently presented
by~\citet{Greene:12}, who found, for a sample of eight massive ETGs,
that metallicity gradients seen within the effective radius continue
to fall to outer radii of $\sim 2.5\,\rm R_e$, while in constrast,
[$\alpha$/Fe] does not drop substantially at large radii in these
objects. Notice that so far, no significant abundance ratio gradients
have been detected in ETGs (e.g.,~\citealt{Kuntschner:10,
  Spolaor:10}).

\section{The effect of galaxy environment}
\label{sec:environment}

\begin{figure*}
\begin{center}
\includegraphics[width=120mm]{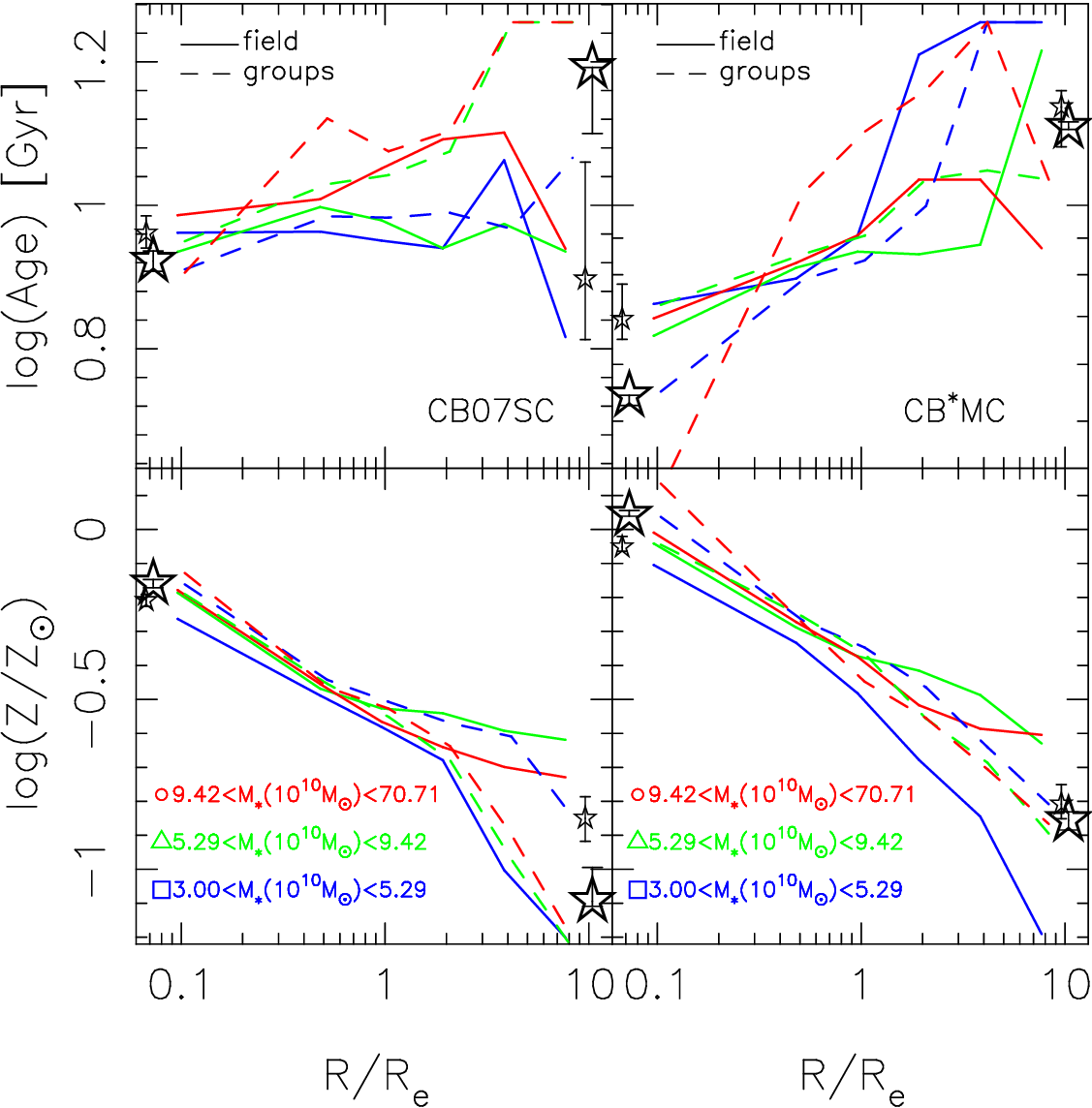}
\end{center}
\caption{Comparison of age (upper panels) and metallicity (lower panels) profiles of
  field (solid curves) and group (dashed curves) ETGs. Left and right panels refer 
to CB07SC and CB$^\ast \!$MC models, respectively (see labels in the lower-right of top panels).
For $\rm R=0.1\,R_e$ ($\rm R = 8\,R_e$), the mean values of
  age and metallicity are plotted with black star symbols on the left
  (right) side of each panel, with small and big stars corresponding
  to field and group galaxies, respectively. Black error bars are the
  mean errors on age and metallicity at $\rm R=0.1\,R_e$  and   $\rm R = 8\,R_e$.}
\label{fig:agemet_profs_env}
\end{figure*}

Fig.~\ref{fig:agemet_profs_env}  compares the  radial profiles  of age
and  metallicity   for  the  subsamples   of  field  and   group  ETGs
(Sec.~\ref{sec:samples}).   For each subsample,  we derive  the median
$\langle g-X\rangle$ colour profiles ($ X=rizYHK$) and fit them, at
the    fiducial   galactocentric    distances,    as   described    in
Sec.~\ref{subsec:color_fitting}.   We show  here only  the  results of
fitting   CB07SC  SSPs,   for  comparison   with  our   previous  work
(e.g.~\citealt[hereafter LF11]{LF:11}), and  those obtained for one of
the  CB$^\ast$ models  (CB$^\ast  \!$MC), as  other  models arrive  at
similar conclusions  (see Sec.~\ref{subsec:SSP_gradients}).  Solid and
dashed  curves  in the  Figure  refer  to  field and  group  galaxies,
respectively.  Black  symbols and  error bars, on  the left  and right
sides of each panel, represent  the mean values of age and metallicity
along with  their error bars at  the extreme fiducial  points of $0.1$
and $8\,\rm R_e$,  respectively. { The mean values  are obtained by
  averaging the values of age  and metallicity for the three mass bins
  at  the   extreme  fiducial  points   of  $0.1$  and   8~$\rm  R_e$,
  respectively.}  For  each subsample, we  derive the inner  and outer
age  (metallicity)  gradients,  \nablati\  and  \dlt\  (\nablazi\  and
\nablazo), as described  in Sec.~\ref{sec:SP_Age_Z}.  Their values are
reported           in           Tab.~\ref{tab:nz_nt_env}           and
Tab.~\ref{tab:nz_nt_env_CB13}, for  CB07SC and CB$^\ast  \!$MC models,
respectively.
\begin{description}
 \item[{\it Central ages.}] In the inner region ($\rm R\le R_e$), group
   ETGs have positive age gradients (age increasing outwards) for each
   mass bin (\nablati$>0$), with the strongest gradients at high mass.
   On the other hand, field ETGs exhibit flatter \nablati's.
   For CB07SC SSPs, field galaxies vary from (slightly
   negative but consistent with) zero gradients at low-mass (\nablati$=-0.01 \pm
   0.06$) to slightly positive at high mass (\nablati$\sim
   0.05$), while for CB$^\ast \!$MC, the \nablati\ is positive, but consistent with zero (at the 
$\sim 1 \, \sigma$ level, see Tab.~\ref{tab:nz_nt_env_CB13}) for all mass bins. 
These findings are consistent with our previous
   results (LF11), based on a larger sample of
   SPIDER ETGs, where  field ETGs (at low- and intermediate-mass) are found with 
   negative -- but small -- age gradients \citep[consistent with,
    e.g.,][]{ferr09}.
\item[{\it Central metallicities.}] No significant difference is detected,
within the errors, between the \nablazi\ of field and group ETGs, although for 
CB$^\ast \!$MC models, at high-mass, the metallicity gradient is significantly steeper for group
(\nablazi$\sim -0.58$) relative to field (\nablazi$\sim -0.37$) ETGs. 
 Note that LF11, based on larger samples of ETGs (i.e. smaller error bars on \nablazi), 
found the inner metallicity gradient of group ETGs to be steeper (at all masses), by 
about $-0.05$, than that of field galaxies. On average, at $0.1\,\rm R_e$, group ETGs are more
  metal-rich than their field counterparts, with $\delta (\log
  Z/Z_\odot) = 0.05 \pm 0.02$ (CB07SC) and $\delta (\log
  Z/Z_\odot) = 0.08 \pm 0.03$ (CB$^\ast \!$MC), averaging over all three mass bins (see
  star symbols in the left side of each bottom panel in
  Fig.~\ref{fig:agemet_profs_env}).
\item[{\it Outer ages.}] In the galaxy outskirts ($\rm R>R_e$), for
  group ETGs, the $Age$ keeps increasing outwards, with very old ages
  at the furthest radial distances probed. Averaging over all three
  mass bins, for group ETGs, we find $\langle$\dlt$\rangle=0.25 \pm
  0.05$~dex (CB07SC) and $\langle$\dlt$\rangle=0.43 \pm 0.09$~dex (CB$^\ast \!$MC). 
On the contrary, for field ETGs, the $Age$ trend is strongly dependent on the stellar population
model one adopts to fit the color profiles. For CB07SC, $Age$ does not increase significantly with 
galactocentric distance, with $\langle$\dlt$\rangle =0.01 \pm 0.07$~dex, while for CB$^\ast \!$MC,
$Age$ increases outwards, with $\langle$\dlt$\rangle =0.21 \pm 0.08$~dex.
Hence, field ETGs have either positive or null age gradients in the outskirts (depending on the model),
while group ETGs have positive age gradients in the outskirts.
\item[{\it  Outer  metallicities.}] For  group  ETGs, the  metallicity
  gradient  is found  to steepen  in the  galaxy outskirts  when using
  CB07SC models ($\langle$\nablazo$\rangle=-0.8  \pm 0.1$, compared to
  $\langle$\nablazi$\rangle=-0.38 \pm 0.02$,  averaging all three mass
  bins).  This result is  somewhat model dependent, being confirmed --
  for  CB$^\ast$   models  --  only  at   low-  and  intermediate-mass
  ($\langle$\nablazo$\rangle=-0.57     \pm    0.07$     compared    to
  $\langle$\nablazi$\rangle=-0.36  \pm  0.03$).  For  field  ETGs,  no
  steepening  is observed  at  intermediate- and  high-mass, for  both
  CB07SC  and CB$^\ast  \!$MC models.   At low-mass,  galaxies  in the
  field  also exhibit  a steeper  gradient in  the outskirts,  for all
  models (\nablazo$\sim  -0.87$ with respect  to \nablazi$\sim -0.32$,
  for CB07SC,  and \nablazi$\sim -0.37$, for  CB$^\ast \!$MC).  Hence,
  the steepening  of the  metallicity gradients in  low-mass ($M_\star
  \sim   4  \times  10^{10}   M_\odot$)  ETGs   is  detected   in  all
  environments,  independent of  the SP  model one  adopts to  fit the
  color     profiles     (consistent     with    what     found     in
  Sec.~\ref{subsec:SSP_gradients}). Notice that at $8\,\rm R_e$, group
  ETGs   are   on   average   less   metal-rich   than   their   field
  counterparts~\footnote{This  is truly  the case  only for  high- and
    intermediate-mass  ETGs. In  fact, as  one can  see in  the bottom
    panels  of  Fig.~\ref{fig:agemet_profs_env},  the  green  and  red
    dashed lines  falls systematically below  the green and  red solid
    lines at $R/R_e > 2$.  At low-mass, group ETGs are more metal-rich
    than their field counterparts, at all masses.}, in contrast to the
  observations in the inner region (see above).


\end{description}


\section{Summary and conclusions}
\label{sec:summary}

We have used optical (SDSS-DR6) and near-infrared (UKIDSS-DR4) data
for 674 massive ETGs to measure, for the first time, colour gradients out to
$\sim 8\,\rm R_{e}$, and study how ETGs assemble their mass. The
sample comprises systems found in galaxy groups and in the field, with
stellar masses ranging from $\sim 3\times 10^{10}\rm M_{\odot}$ to
$\sim 7\times 10^{11}\rm M_{\odot}$. The critical systematic effects
that may affect our conclusions have been carefully
taken into account, and we show that none of them (e.g. sky
subtraction, colour dependence of the PSF outer wings, and the
approach adopted to derive the colour profiles) introduce
any systematic biases.
The stacked colour profiles show remarkable linearity out to $\rm R
\sim 8\,\rm R_{e}$, getting bluer in the outer regions, irrespective of
the mass bin, with the exception of $g - r$ where a curvature is apparent.
Nevertheless, this effect is small and does not affect the stellar population
analysis, as the deviations are comparable to the random
errors at the radii where the curvature is detected.
For the most massive galaxies, the stacked $g-r$ colour profile is
compared with that recently obtained by~\citet{TalvanDokkum:11},
finding good agreement.

Colour   profiles  are  translated   into  SSP-equivalent   $Age$  and
metallicity gradients  using a  variety of stellar  population models,
assuming either  a Chabrier  or Salpeter IMF,  as detailed  in Section
5.2.   In particular,  we exploit  the new,  state-of-the-art, stellar
population models of Charlot  \& Bruzual (2013, in preparation), based
on different  stellar libraries  (IndoUS, Miles, STELIB,  BaSeL). {
  Errors  on  both age  and  metallicity  are  carefully estimated  by
  accounting  for uncertainties  on  observed colors.}   We find  very
unambiguously  that metallicity  decreases from  $0.1$ out  to $8\,\rm
R_{e}$ (negative gradient), regardless of the mass bin considered (see
Tab.~6). This result is  independent of the adopted stellar population
model and  IMF.  In  addition, we find  that the  metallicity gradient
tends to steepen at  large galacto-centric radii.  The significance of
this  result changes with  the adopted  stellar population  models and
galaxy  mass bin (being  more significant  at low  mass).  As  for the
$Age$ parameter, the situation is more complex.  From $0.1$ to $1\,\rm
R_{e}$,  we find  that  for intermediate-  and  high-mass ETGs,  $Age$
increases  with  radius,  while  at  low mass,  no  clear  trend  with
galacto-centric  distance is  detected, consistent  with  our previous
results  (see e.g.~\citealt{LF:11} and  references therein).   The new
result of the  present work is that in the outer  regions, from $1$ to
$8\,\rm R_{e}$, the stellar populations of ETGs are even older than in
their centres.  This  is more significant for galaxies  in the highest
stellar mass bin (i.e.  $M_\star\simgt 10^{11}M_\odot$).
We emphasize that using models with extended star formation histories,
like exponentially declining ($\tau$) and finite burst models, instead
of SSPs, yields similar results within the quoted error bars. We also
confirm that the presence of uniformly distributed internal reddening
(dust) does not affect the trends of age and metallicity with
galactocentric distance.

Finally, we analyze how the age and metallicity profiles depend on the
environment  where  galaxies reside.   Group  ETGs  have positive  age
gradients out to $\sim 8\,\rm R_e$, with a very old stellar population
in  the outskirts.   Their metallicity  gradients steepen  at  $\rm R>
1-2\,\rm R_e$.  In  constrast, field ETGs do not  have significant age
gradients in the inner region,  while at $\rm R\simgt 1\,\rm R_e$, the
$Age$ shows  either a flat  trend or increases outwards  (depending on
the  model).  The  metallicity  gradient  of field  ETGs  is found  to
steepen  significantly in  the  outskirts only  for low-mass  systems.
Notice that the results for group galaxies are essentially the same as
those for  the whole sample of  ETGs (see above),  consistent with the
fact  that most  of our  sample  ($\sim 80\%$)  comprises ETGs  either
residing in (or close  to) galaxy groups (see Sec.~\ref{sec:samples}).
Our  findings  point  to  the  importance  of  aperture  effects  when
comparing  the  stellar  population   content  of  ETGs  in  different
environments, as in the galaxy outskirts, group galaxies are older (or
coeval, depending on  the model) and less metal-rich  than field ETGs,
while the opposite holds in the central regions, { where group ETGs
  are  found to  be more  metal-rich than  their  field counterparts}.
This might explain, at least  in part, why different results have been
reported in  the literature about the environmental  dependence of the
metal  content  of  ETGs  (see  Paper  III),  with  different  studies
reporting   field  ETGs   to   be  more   metal-rich~\citep{Thomas:05,
  deLaRosa:07,    Kunt:02,   Clemens:09,   Zhu:10},    as   metal-rich
as~\citep{BERN:06,  Annibali:07}, or even  more metal-poor  than their
cluster counterparts~\citep{GALL:06}.

{ We have  shown here that the stellar populations  in the halos of
  ETGs are consistently older and more metal poor than in their cores.
  This trend depends on mass  and, to a lower degree, environment, but
  overall, the  age and metallicity  radial gradients are  similar for
  all  ETGs   in  the  sample   (i.e.   more  massive   than  $3\times
  10^{10}$M$_\odot$).  Within  the core of the  galaxies, the observed
  metallicity gradients  -- between \nablazi$=-0.3$ and  $-0.4$ -- are
  consistent  with  the simulations  of  \citet{koba04} for  non-major
  merger  systems.  For galaxies  where major-merging  is significant,
  her simulations give shallower metallicity gradients, around $-0.2$.
  It  is worth  noting  that our  metallicity  gradients get  slightly
  steeper with  increasing mass, supporting  the scenario of  fast and
  early formation  for the most massive ETGs.   Extending the analysis
  radially  out to  $8{\rm R}_e$  allows us  to explore  the different
  channels of  star formation and  assembly in ETGs.  Our  results are
  consistent with the scenario  of \citet{oser10}, where two phases of
  formation  operate:  an early-phase  of  in-situ  star formation  --
  building  the core of  the galaxy  -- followed  by the  accretion of
  small  satellites.   The  recent simulations  of  \citet{Lackner:12}
  quantify  in  more detail  the  buildup  of  the stellar  component,
  concluding that major mergers  do not dominate the accretion history
  of massive  galaxies, with smaller systems consisting  of older, and
  more  metal-poor stellar  populations  building up  the outer  halo,
  consistent  with  our   analysis.   Their  simulations  suggest  age
  (metallicity)  differences of  $\Delta  t\sim 2.5$~Gyr  ($\Delta\log
  Z\sim -0.15$\,dex) between accreted  stars and those formed in-situ.
  This  process is  fully consistent  with the  observed  evolution of
  massive  early-type  galaxies  on  the mass-size  plane  \citep[see,
    e.g.][]{daddi05,truj06,truj11,vdk08},  where the  core  is quickly
  formed during  an early phase  \citep[see, e.g.][]{ig:09b}, followed
  by   the  growth  of   the  halo   via  minor   mergers  \citep[see,
    e.g.][]{naab09}.  }

\appendix

\section{Comparison of parametric and non-parametric light profiles}
\label{sec:light_profiles}

\begin{figure*}
\begin{center}
\includegraphics[width=120mm]{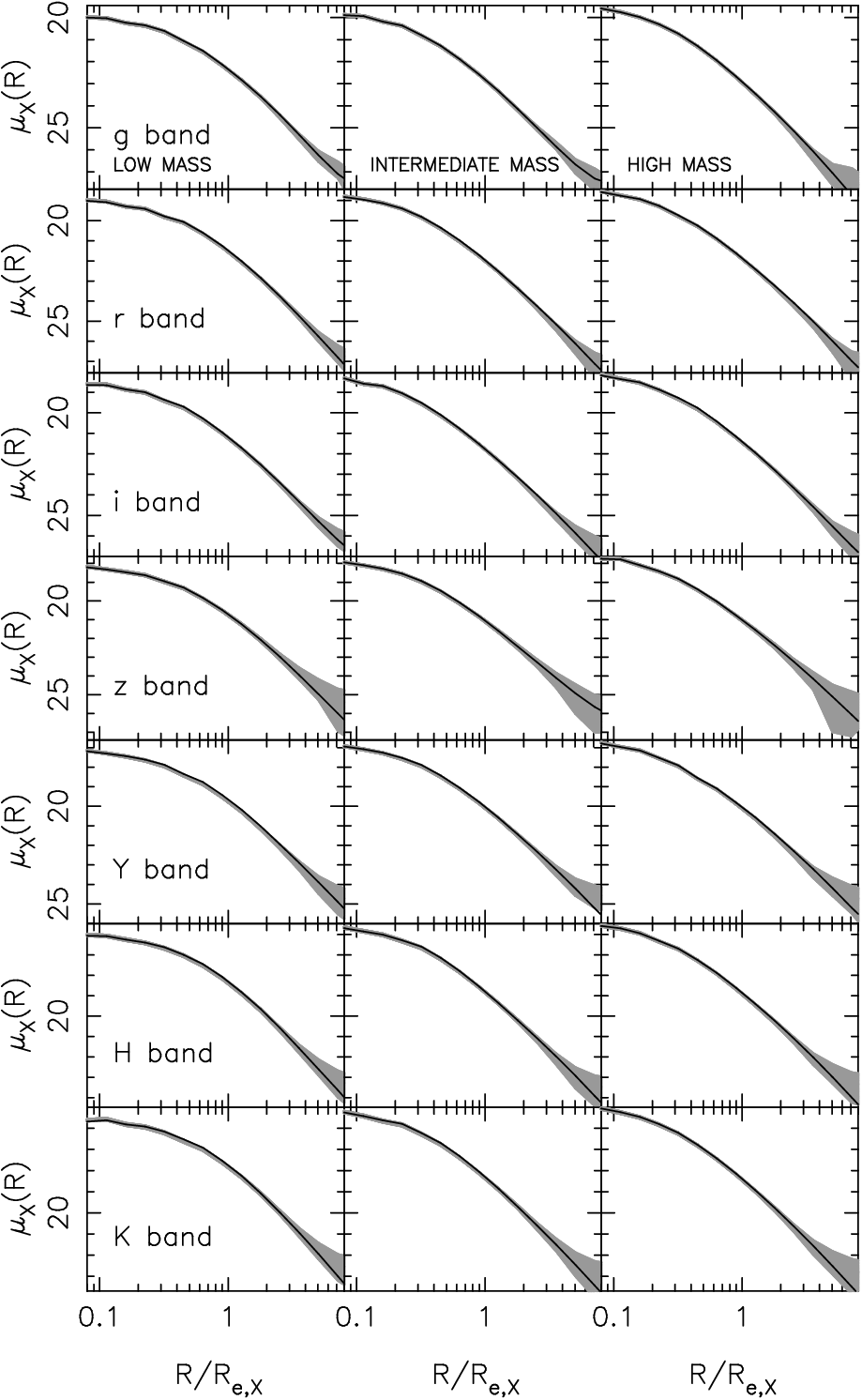}
\end{center}
\caption{Comparison  of parametric  and non-parametric  light profiles
  for   low-   (left-),   intermediate-   (middle-),   and   high-mass
  (right-panels) ETGs,  and each waveband out of  $X=rizYHK$ (from top
  to  bottom,  as labelled  in  left  panels).   Black lines  are  the
  parametric  profiles, while  grey regions  mark the  1~$\sigma$ mean
  confidence  intervals around  non-parametric profiles.   Notice that
  black curves  fall in the  grey regions for  all mass bins,  and all
  wavebands, supporting our approach to estimate colour profiles. }
\label{fig:light_profiles}
\end{figure*}

{    Fig.~\ref{fig:light_profiles}    compares    parametric    and
  non-parametric stacked  light profiles for each of  the three galaxy
  mass bins (from left to right), and each waveband, $X=grizYHK$ (from
  top to  bottom), used in  the present study.   We do not  consider J
  band, as this  is not used for the  stellar population analysis (see
  Sec.~\ref{sec:col_prof_mass} for  details).  For each  galaxy image,
  and a given waveband $X$,  we measure mean surface brightness values
  on a  set of  concentric circles, with  radii equally spaced  by 0.5
  pixels,  out to  a maximum  distance of  $\sim 110$~arcsec  from the
  galaxy center.  All  objects around the given galaxy  are masked out
  with  the  software  2DPHOT,  as described  in~\citet{LBdC08}.   The
  surface  brightness  profiles of  all  galaxies  are  scaled by  the
  corresponding  $R_{e,X}$'s, and  normalized  to have  the same  flux
  (within an aperture of  $2R_{e,X}$).  The so-normalized profiles are
  averaged  within radial bins  logarithmically spaced  by $0.15$~dex.
  This procedure provides non-parametric light profiles (not corrected
  for PSF), with grey regions in Fig.~\ref{fig:light_profiles} marking
  the  1~$\sigma$  mean confidence  contours  around  them.  The  same
  procedure, applied to the PSF-convolved best-fitting two-dimensional
  S\'ersic  models, gives  the parametric  light profiles,  plotted as
  black curves  in Fig.~\ref{fig:light_profiles}.  Overall,  we find a
  good  agreement between  observed  and model  profiles,  out to  the
  largest  galactocentric  distances considered  in  this work  ($\sim
  8$~$\rm R_e$). This supports  previous claims that a single S\'ersic
  law is  able to describe  accurately the light distribution  of ETGs
  out  to $\sim  8 \,  \rm  R_e$~\citep{Kormendy:09, TalvanDokkum:11},
  supporting the robustness of  our parametric colour estimates over a
  wide radial range. }

\section*{Acknowledgments}

{  We thank  the anonymous  referee  for the  helpful comments  and
  suggestions, that helped us  to improve the present manuscript.}  We
have    used    data   from    the    Sloan    Digital   Sky    Survey
(http://www.sdss.org/collaboration/credits.html).   IGR acknowledges a
grant from the Spanish Secretar\'\i  a General de Universidades of the
Ministry  of Education, in  the frame  of its  program to  promote the
mobility of  Spanish researchers to foreign  centres.  GB acknowledges
support from  the National  Autonomous University of  M\'exico through
grants IA102311  and IB102212.   We have used  data from the  4th data
release  of the  UKIDSS survey  \citep{Law07}, which  is  described in
detail in  \citet{War07}.  Funding for  the SDSS and SDSS-II  has been
provided  by  the  Alfred  P.   Sloan  Foundation,  the  Participating
Institutions, the National Science Foundation, the U.S.  Department of
Energy,  the  National   Aeronautics  and  Space  Administration,  the
Japanese  Monbukagakusho,  the  Max  Planck Society,  and  the  Higher
Education Funding Council for England.



\label{lastpage}

\end{document}